\documentclass[10pt]{iopart}
\usepackage{epsfig,graphicx,graphics}

\newcommand{\be}{\begin{equation}}
\newcommand{\ee}{\end{equation}}
\newcommand{\bea}{\begin{eqnarray}}
\newcommand{\eea}{\end{eqnarray}}
\newcommand{\order}{{\mathcal{O}}}
\newcommand{\bxi}{{\mbox{\boldmath $\xi$}}}
\newcommand{\bsigma}{{\mbox{\boldmath $\sigma$}}}

\newcommand{\bm}{{\bf m}}

\newcommand{\bra}{{\langle}}
\newcommand{\ket}{{\rangle}}

\newcommand{\bp}{{\bf p}}

\newcommand{\half}{{\frac{1}{2}}}
\newcommand{\atanh}{{\rm atanh}}

\newcommand{\mcirc}{\circle*{2}}
\newcommand{\bcirc}{\circle*{7}}
\newcommand{\Bcirc}{\circle{24}}

\begin{document}

\title[The role of the T-helper/T-suppressor ratio
in the adaptive immune response]{\bf The role of the T-helper/T-suppressor ratio
in the adaptive immune response: a dynamical model}
\author{A Annibale$^\dag$, LA Dziobek-Garrett, H Tari}
\address{$\dag$ Department of Mathematics, King's College London, The Strand,
London WC2R 2LS, United Kingdom}

\begin{abstract}
Recent experimental studies have suggested the ratio between T-helper 
and T-suppressor lymphocytes as an index 
of immunosuppression in HIV, cancer, immunosenescence and inflammatory and auto-immune diseases.
However, a quantitative understanding of the impact of this ratio on the immune response
has lagged behind data and its validity 
as a tool for prognostic monitoring or therapeutic target remains an open question.
In this work, we use statistical physics and dynamical systems 
approaches to analyze the time-dependent response 
to an antigen, of a simplified model of the
adaptive immune system, which 
comprises B, T-helper and T-suppressor lymphocytes. 
The model is remarkably robust against changes in the noise level and 
kinetic parameters, but it is very sensitive to changes 
in the ratio between T-helper and T-suppressor 
lymphocytes, exhibiting, in particular, a transition from a responsive 
to an immuno-suppressed phase, as the ratio is lowered below a critical 
value, which is in line with experiments.
This result supports the validity of the T-helper/T-suppressor ratio as an index of immunosuppression 
and may provide a useful theoretical benchmark to interpret and compare experiments.
\end{abstract}

\section{Introduction} 
Recent years have seen a surge of mathematical and computational approaches aimed at modelling the immune 
system from a systemic perspective.
Many popular models have been formulated in terms of ordinary
differential equations
\cite{Nowak00, Leon03, Burroughs06, FouchetRegoes08, self_tolerance, DeBoer17, Kim07, CarLeoCar07}
and partial
differential equations \cite{Antia03, OnsumRao07, Carneiro06}, as these provide intuitive frameworks to understand 
the dynamics of average quantities (e.g. cell concentrations). 
However, they normally ignore microscopic details and stochasticity, 
due to noise in the biological environment or 
fluctuations in cellular densities, and generally require the
estimation of a large number of unknown parameters. 
Agent-based simulations 
\cite{agents1, agents2, Scherer06, Casal05} and machine learning approaches
\cite{max-ent1, max-ent2, bayes, Becca16} 
have been successful in incorporating statistical noise and 
microscopic information 
(e.g. cellular interactions, antibodies sequences etc.),  
however, they usually require more significant computational efforts.
In recent years, statistical mechanical approaches
\cite{Parisi, Adriano_stat_mech, AB, saturation, medium, Silvia_3} have supplied useful models to handle the immune system's intricate patterns of interactions while 
remaining analytically tractable. 
However, most of these studies are performed at equilibrium: 
statistical mechanical models looking at the dynamics of the immune system after antigenic stimulus are still lacking or are at their infancy 
\cite{Adriano_stoch_dyn, Silvia_2}. One of the aims of this work is to advance non-equilibrium statistical mechanical models of the adaptive immune system.

The adaptive immune system is a complex network of cells that work together 
to defend the body against pathogens such as bacteria, virus or tumor cells.  
The major agents of adaptive immunity are lymphocytes, which are 
broadly divided into T and B cells.
Each B cell is equipped with a large number 
$n=\order{(10^5)}$ of identical B cell receptors 
(BCR) on its surface, 
which are able to bind to specific pathogens. 
When a B cell comes across a pathogen that can bind its receptors,
it ingests it, partially degrades it, and exports fragments of it, i.e. 
antigens, to the cell surface, where 
they are presented in association with proteins known as MHC molecules.
T cells are also equipped with receptors (TCR), but in contrast to B 
cells, they do not bind to the antigen directly, but to the MHC-antigen complex on the surface of 
antigen presenting cells,
such as B cells, dendritic cells and other immune cells. 
T cells can be divided into CD4+ (helper) and CD8+ (cytotoxic/suppressor)
cells, which bind different classes of MHC-molecules. 
CD4+ cells bind to class II MHC-molecules and have the role of signaling B 
cells to initiate an immune response. 
CD8+ cells bind to class I MHC-molecules and can be divided 
into cytotoxic T cells, which destroy
infected or cancer cells, and suppressor T cells, which switch off the 
immune response. 
In healthy hosts, the ratio between CD4+ and CD8+ cells 
is normally above one, while low or inverted CD4+/CD8+ ratios have been associated with 
impaired immune function in
inflammation and autoimmune diseases \cite{inflamatory_1,inflamatory_2,inflamatory_3,inflamatory_4}, 
cancer \cite{cancer_1,cancer_2} and
immunosenescence \cite{age_1,age_2,age_3}. 
Furthermore, recent studies have suggested that the CD4+/CD8+ ratio may affect
the progression of HIV infections and response to antiretroviral therapies 
\cite{HIV_1,HIV_2,HIV_3,HIV_4,HIV_5}, and may be a marker for viral reservoirs 
\cite{reservoir_1,reservoir_2,reservoir_3,reservoir_4} in HIV-positive patients. However,  
a sturdy relation between CD4+/CD8+ ratios 
and viral reservoirs has not been proven and the 
general validity of the CD4+/CD8+ ratio as a tool for prognostic monitoring or therapeutic target 
remains an open question \cite{opinion}. 

With the advance of molecular immunology, the existence of several 
subtypes of CD4+ and CD8+ lymphocytes
has been documented and a subtype of CD4+ cells
functioning as suppressor cells has been uncovered
\cite{CD4+CD25_90, CD4+CD25_95}. These were named T-regulatory cells, and their 
biomarkers have been 
identified recently after a long-standing debate \cite{CD4+CD25Foxp3} and localized in the cell \cite{localized}. 
Similarly to the CD4+/CD8+ ratio, the T-helper/T-regulatory cell imbalance has been suggested as an index of 
immunosuppression in cancer patients 
\cite{cancer_2, ratio_cancer, ratio_cancer_2}, HIV patients
\cite{ratio_HIV_review,Treg_HIV_2015,ratio_HIV}, immunosenescence \cite{ratio_age} and in inflammation 
diseases \cite{ratio_inflamatory}. Results generally suggest that lower values of 
T-helper relative to T-regulatory cells are associated with 
unfavourable prognosis. However, different studies have used
different biomarkers to identify T-regulatory cells 
(resulting in populations with different degrees of enrichment in 
T-regulatory cells)
and have focused on different populations of T-helper cells (sometimes
all CD4+ cells, other times only subtypes of them
). In addition,
cell imbalance has been measured in different ways 
(sometimes in terms of differences, others in terms of ratios 
between cell concentrations), and is subject to fluctuations in T cell counts, 
meaning that a global consensus on its reliability 
as a biomarker for immunosuppression has not been reached.

To add complexity, the CD8+/T-regulatory ratio has also been
associated with clinical outcome in cancer recently \cite{CD8+Treg}, 
and with the increasing number of cell subtypes being uncovered, 
many different combinations of cell subtypes can be considered. 
However, assessing the significance 
of any observed correlation between the imbalance of different combinations of 
cell subtypes
and prognosis 
in a particular disease,
based on data from a limited number of patients,  
cannot be done reliably without a suitable theoretical model. 

The aim of this work is to provide a theoretical framework to understand the impact of the 
T-helper/T-suppressor ratio on the response of the immune system to an antigen. 
We will move away from a detailed characterization of T-helper and T-suppressor cells in terms of biomarkers, 
and we simply 
assume that there are two broad categories of T cells: T-helper, 
that activate B cells, and T-suppressor, 
that inhibit B cells, in more or less direct ways, including cytotoxic and 
regulatory activity. Although our definition 
of T-suppressor includes different cell lineages i.e. CD8+ cytotoxic, CD8+ suppressor and CD4+ regulatory, 
we note that 
CD4+ regulatory cells are present in 
normal peripheral blood in low numbers, typically 5-10\% of CD4+
T cells \cite{proportion_Treg}, so 
the T-helper/T-suppressor ratio considered in this work, does not deviate significantly from the CD4+/CD8+ ratio 
considered in the literature.
This paper will articulate as follows:
In Sec. \ref{sec:bio} we provide a 
brief description of the adaptive immune system and summarize the main 
cellular reactions involved in its response to an antigen. 
In Sec. \ref{sec:model} we define a network model of the adaptive immune system,
comprising T-helper, T-suppressor and B cells,  
and study its response to a single antigen, 
for a large range of its control parameters, in the thermodynamic limit. 
In Sec. \ref{sec:Gillespie} we simulate the reactions between immune cells 
using a Gillespie algorithm and check the accuracy of our model. In Sec. 
\ref{sec:extensions} we look at possible generalizations and extensions of our model. Finally, in Sec. \ref{sec:conclusions} we summarize our results and propose pathways for future 
work.

\section{The adaptive immune system: a brief description}
\label{sec:bio}
T-helper cells get activated upon binding antigen presenting cells, like B cells, dendritic cells or other immune 
cells presenting antigens on their surface. 
When active, T-helper cells proliferate and release proteins called cytokines, which activate B 
cells. 
Active B cells undergo clonal expansion, i.e. form many copies of identical cells sharing 
the same antigen receptors, and secrete antibodies, 
i.e. a free form of those receptors, that can recognize and neutralize the antigen \cite{Abbas}. 
Clonal expansion usually involves migration of B cells to the germinal 
centre\footnote{Germinal centres are transient structures that
form within peripheral lymphoid organs in response to
T cell-dependent antigen}, where B cells proliferate at a rate that is
unparalleled in mammalian tissues \cite{germinal}. Random mutations during clonal expansion cause the production of 
antibodies with a broad range of binding affinities for their antigen. 
B cells with unfavourable mutations will not get sufficiently activated by the antigen and T-helper cells 
and will die, while those with improved affinity will be stimulated to clone themselves.
This leads to an effective 
selection processes within the germinal centre, known as affinity maturation, which results in 
the production of antigen-specific B cells within one week, in mammalians \cite{B_GC}.

T-suppressor cells switch off the immune response. 
Although the mechanics of regulatory T cells 
is less clear than that of cytotoxic T cells 
\cite{howTregWork, TregActivation_transplant_autoimmunity}, their ability to directly 
suppress B cells has been well-documented in recent studies 
\cite{Zhao06, Vignali08, Wang13, Gotot12, Lim05, Iikuni09}. 
From a modelling perspective, we will not distinguish between 
cytotoxic and regulatory cells, and we will simply regard them as 
T-suppressor cells, with the ability to 
reduce the number of effector \footnote{Effectors B cells are B cells that secrete large volumes of antibodies.} B cells, 
by either killing them or inhibiting their production of antibodies.

Below, we list the reactions that we shall include in our model. 
These occur in different parts of the organism, e.g. T and (antigen 
presenting) B cells mostly 
interact in the lymph nodes (organs of the lymphatic system), 
while lymphocytes and antigens also interact in infected tissues and other parts of the lymphatic system.
In this study, we will not consider spatial effects and we will regard all the reactions below as 
taking place in the broad lymphatic system, which communicates lymphatic organs (such as lymph nodes) 
with infected tissues and with the vascular system. 
For simplicity, we suppose that each pathogen has a 
single epitope, that we will loosely call antigen. We have:
\begin{itemize}
\item Pathogens replicate in the host at rate $r$, leading to proliferation of antigens 
\be
Ag \rightarrow^{\!\!\!\!\!\! r} \,2 Ag
\label{eq:Ag_r}
\ee
\item B cells bind to antigens, thus becoming antigen presenting B cells (APB),
at a rate $\pi^+$, which depends on the 
affinity between their receptors and the antigen
\be
Ag+B \rightarrow^{\!\!\!\!\!\! \pi^+} APB
\label{eq:APB}
\ee
\item T cells get activated when they bind an APB, at a rate $W$
\be
T+APB \rightarrow^{\!\!\!\!\!\! W} T^\star + APB
\label{eq:T_active}
\ee
We note that this is not the only mechanism for T cell activation, as the 
latter can also be driven by 
dendritic cells. We will show in Sec. \ref{sec:model} 
that the inclusion of dendritic cells has only a small quantitative effect on the 
model, hence, for simplicity, we will mostly negect it.
\item Antigen presentation is a reversible process, that can be switched off 
\cite{Lankar02}, 
at a rate $\pi^-$ 
\be
APB \rightarrow^{\!\!\!\!\!\! \pi^-} B
\label{eq:APB_remove}
\ee
Given that B cell activation requires antigen degradation,
we will assume that antigen fragments freed up are not able to replicate, and do 
not contribute to the population of replicating antigens.
\item Activated T cells induce B clonal expansion 
\bea
T^\star+B \rightarrow^{\!\!\!\!\!\! \lambda^+} T^\star  + 2B
\label{eq:B_expand}
\eea
or contraction
(via e.g. cytolysis \cite{Zhao06, Vignali08}),
\bea
\quad T^\star+B \rightarrow^{\!\!\!\!\!\! \lambda^-} T^\star  
\label{eq:B_contract}
\eea
depending on the nature, helper or suppressor, of the T cell (we will introduce a binary variable 
to discern helpers and suppressors in the next section). We 
denoted $\lambda^+$ and $\lambda^-$, the 
rates of clonal expansion and contraction, respectively. 
\item B cells are kept in a state of activated apoptosis while
undergoing clonal expansion in the germinal centre and compete 
for survival signal from T-helper cells \cite{double_sig1}. We assume that 
when two B cells compete for the same resources, one will be selected
at a rate $\delta$
\be
B+B \to^{\!\!\!\!\!\! \delta}  B.
\label{eq:B_compete}
\ee
\end{itemize}
To keep the model simple, we will neglect several other 
processes such as antigen mutation, clonal expansion of 
T cells, activation of macrophages 
and other types of immune cells, production of antibodies 
by B cells, and affinity maturation of B cells in the germinal centre. 
We will simply assume that, due to the latter  
process, the affinity $\pi^+$ between BCR and antigen 
is an increasing function of time. 

Despite the many simplifying assumptions, we believe that the above reactions capture the basic 
principles of an immune response to an antigen. In the next section, we will use them to build an analytically tractable model that links the microscopic dynamics of cellular activation 
with the macroscopic dynamics of clonal concentrations. 
We will assume that T and antigen presenting B cells interact via a network, while 
all the other species are well-mixed within the lymphatic system.
In contrast to more traditional 
phenomenological approaches formulated in terms of differential equations, the resulting model 
will incorporate information on microscopic interactions and
stochasticity, and will be formulated in terms 
of a minimal number of equations (i.e. four), which do not require the estimation of a large number of unknown
parameters and are extremely robust to parameters variation. 
Results will highlight the role played by the ratio between T-helper and T-suppressor lymphocytes on immunosuppression, 
and will expose the role played by stochasticity on homeostasis. Both the T-helper/T-suppressor ratio and the mechanics of homeostasis have been the subject of 
intensive experimental investigation in recent years.

\section{A network model of the adaptive immune system}
\label{sec:model}
We consider a population of T cells, each labelled by 
$i=1,\ldots,N$, and a population of B clones\footnote{A B clone is the ensemble of all B cells that have the same receptors and  
thus respond to the same antigen.}, each
labelled by $\mu=1,\ldots,P$. 
Experimental lymphocytes counts in humans
estimate the total number of T {\it cells} to be $\order{(10^{11})}$ \cite{T_cells_10_to_11}, 
the total number of B {\it cells} to be roughly of the same order \cite{B_cells_10_to_10-11}, and 
the number of B {\it clones} to be $\order{(10^7)}$ or higher \cite{B_clone_size}.
Hence, we set the number of B {\it cells} to be $\phi N$ with $\phi=\order{(1)}$ and the number of B 
{\it clones} to be sub-extensive in $N$, i.e. $P=\alpha N^\gamma$ 
with $\gamma\in(0,1)$ and $\alpha=\order{(1)}$, so that each B clone contains, 
on average, $N^{1-\gamma}\phi/\alpha$ cells. 
\footnote{Clone sizes are known to be heterogeneous 
\cite{Walczak}, here the average over all clones is given.}
For simplicity, we ignore cross-reactivity effects between B clones and 
antigens and we assume that each B clone $\mu$ 
is able to bind only to one type of antigen (assumed to have a single epitope)
also labelled by $\mu$.

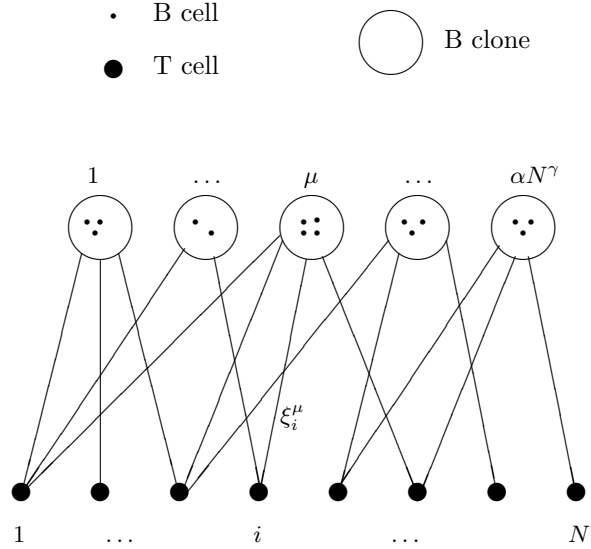
\begin{figure}
\hspace*{2cm}
\begin{picture}(100,165)
\thinlines
\put(65,150){\mcirc}\put(80,148){B cell}
\put(65,130){\bcirc}\put(80,128){T cell}
\put(170,140){\Bcirc}\put(190,138){B clone}
\put(55,87){\small $1$}
\put(55,72){\mcirc}
\put(60,72){\mcirc}
\put(58,68){\mcirc}
\put(95,87){\small $\ldots$}
\put(96,72){\mcirc}
\put(102,68){\mcirc}
\put(137,87){\small $\mu$}
\put(137,72){\mcirc}
\put(142,73){\mcirc}
\put(137,68){\mcirc}
\put(142,68){\mcirc}
\put(175,87){\small $\ldots$}
\put(175,72){\mcirc}
\put(178,68){\mcirc}
\put(182,72){\mcirc}
\put(215,87){\small $\alpha N^\gamma$}
\put(217,72){\mcirc}
\put(223,72){\mcirc}
\put(220,68){\mcirc}

\put(27,-49){\small $1$}
\put(62,-49){\small $\ldots$}
\put(118,-49){\small $i$}
\put(128,-4){\small $\xi_i^\mu$}
\put(170,-49){\small $\ldots$}
\put(237,-49){\small $N$}
\multiput(60,70)(40,0){5}{\Bcirc}
\multiput(30,-30)(30,0){8}{\bcirc}
\put(53,60){\line(-1,-4){22}} 
\put(60,58){\line(0,-4){91}}
\put(67,60){\line(1,-4){22}} 
\put(92,62){\line(-2,-3){62}} 
\put(103,59){\line(1,-5){18}} 
\put(169,65){\line(-4,-5){76}}
\put(173,60){\line(-1,-4){23}} 
\put(138,58){\line(-1,-5){18}}
\put(144,59){\line(2,-5){36}} 
\put(129,65){\line(-2,-5){38}}
\put(128,67){\line(-1,-1){96}}
\put(191,65){\line(1,-5){19}} 
\put(222,59){\line(1,-5){18}}
\put(211,63){\line(-2,-3){63}} \put(217,59){\line(-2,-5){37}}
\end{picture}
\vspace*{1.5cm}
\caption{Schematic representation of the interactions between T cells and 
B clones. A link $\xi_i^\mu$ between T cell $i$ and B clone $\mu$ means that 
T cell $i$ can bind to B cells in clone $\mu$. The number of T cells 
signaling a clone $\mu$ is of the same order as the number of B cells in 
clone $\mu$, assumed $\order{(N^{1-\gamma})}$. In the figure, they 
are precisely the same and $\order{(1)}$ and are meant for illustration only.}
\label{fig:xi_mu}
\end{figure}

Next, we define the statistics of the interactions between T cells and B clones (see Fig. \ref{fig:xi_mu} for a schematic representation).
We introduce a binary variable $\xi_i^\mu\in\{1,0\}$ to indicate whether 
($1$) or not ($0$) T cell $i$ can bind cells in B clone $\mu$. 
We regard the variables $\{\xi_i^\mu\}$ as random i.i.d., with distribution
\be
p(\xi_i^\mu)=\frac{c_\mu}{N^{\gamma}}\delta_{\xi_i^\mu,1}+ \left(
1-\frac{c_\mu}{N^{\gamma}}
\right)\delta_{\xi_i^\mu,0}
\label{eq:p_zeta}
\ee 
where $c_\mu=\order{(N^0)}~\forall~\mu$, so that 
each B clone $\mu$ has, on average, 
$\bra \sum_{i=1}^N \xi_i^\mu \ket=c_\mu N^{1-\gamma}$ conjugate T cells. 
If all the T cells that react with a B clone belong to the same T clonotype, 
then $c_\mu N^{1-\gamma}$ is the average size of the T 
clonotype conjugate to B clone $\mu$, at rest. However, 
this assumption is not necessary and we admit the possibility 
that a B clone interacts with different T clonotypes. In this case the 
$c_\mu N^{1-\gamma}$ T cells interacting with B clone $\mu$ may belong to 
different T clonotypes.   
In either case, choice (\ref{eq:p_zeta}) corresponds to the scenario 
where the number of T cells that can signal a B clone is of the 
same order as the number of B cells in that clone.
Biologically, this would 
seem the most plausible scaling, as it avoids, on the one hand, 
a redundancy of B cells 
that would not get sufficiently signaled by 
a comparatively small number of conjugate T cells, 
and on the other hand, a waste of T 
cells in signaling a comparatively small number of B cells. 
%

We denote with $\psi_\mu$, $b_\mu$ and $p_\mu$, the population densities of  
antigens of type $\mu$, B clone $\mu$ and APB of type $\mu$, respectively.
It is convenient to define clonal densities as the number of cells in a clone 
divided by the average resting number of conjugate T cells, i.e.
$b_\mu\!=\!B_\mu/c_\mu N^{1-\gamma}$, where $B_\mu$ is the 
number of cells in B clone $\mu$ and similarly 
for $p_\mu\!=\!P_\mu/c_\mu N^{1-\gamma}$ 
and $\psi_\mu\!=\!\Psi_\mu/c_\mu N^{1-\gamma}$. 
We can write the following equations for
reactions (\ref{eq:Ag_r}), (\ref{eq:APB}) and (\ref{eq:APB_remove})
\bea
\frac{d}{dt} p_\mu&=&\pi_\mu^+ \psi_\mu b_\mu-\pi_\mu^- p_\mu
\label{eq:p_dyn}
\\
\frac{d}{dt} \psi_\mu&=&r_\mu \psi_\mu-\pi_\mu^+ \psi_\mu b_\mu
\label{eq:psi_dyn}
\eea
with $p_\mu, b_\mu$ and $\psi_\mu$ dimensionless and 
the kinetic coefficients 
$\pi_\mu^+, \pi_\mu^-, r_\mu$ having the unit of inverse time.
In the absence of B cells, (\ref{eq:psi_dyn}) gives 
$r_\mu=\ln 2/t_\mu^\star$, where $t_\mu^\star$ is the doubling time 
of antigen $\mu$. Doubling times vary across different diseases, 
however, typical values in human hosts, at the early stage of an infection, 
when an immune response has not started yet, are estimated 
to range between a few hours (bacteria) and a day (virus), see e.g. 
\cite{Ag_doubling_b, Ag_doubling_v}.
It is then convenient 
to measure time in days, so to work with $\order{(1)}$ replication rates. 

T cells can be helpers or suppressors. 
For each T cell $i$, we introduce a variable $\eta_i$ which takes values 
$1$ if $i$ is helper and $-1$ if it is suppressor. 
We assume each $\eta_i$ to be identically and independently sampled from 
\be
P(\eta)=\frac{1+\epsilon}{2}\delta_{\eta,1}+\frac{1-\epsilon}{2}\delta_{\eta,-1}
\label{eq:h_ratio}
\ee
The parameter $-1< \epsilon< 1$ quantifies the imbalance 
between T-helper ($\eta=1$) and T-suppressor ($\eta=-1$) cells
and is directly related to the T-helper/T-suppressor ratio $R$,
measured in experiments
\be
R=\frac{1+\epsilon}{1-\epsilon}.
\label{eq:R}
\ee
For $\epsilon=0$, T-helper and T-suppressors cells 
are present in equal proportions, both equal to $1/2$ 
(i.e. $R=1$), while for $\epsilon>0$ ($\epsilon<0$) T-helper cells are more (less) than T-suppressor cells.

We represent the state of each T cell $i$ (active or inactive) with a variable
$\sigma_i$ which takes value $1$ if $i$ is active and $0$ otherwise. 
Helper T cells get activated via reaction (\ref{eq:T_active}), i.e.
when their receptors bind to an APB. For simplicity, we assume that T cells update their 
state at regular time intervals of duration $\Delta$, according to the stochastic rule
\be
\sigma_i(t+\Delta)=\theta(\sum_{\mu=1}^P \xi_i^\mu p_\mu(t)-z(t))
\label{eq:sigma}
\ee
where 
$\theta(x)=0$ for $x\leq 0$ 
and $\theta(x)=1$ for $x>0$, and 
$z(t)$ is a zero-averaged random variable with suitably normalised variance,
drawn, at each time $t$, from a symmetric distribution $p(z)$,
which mimicks "fast" noise in the biological environment or stochasticity in T cell activation.
In the absence of noise ($z=0$), equation (\ref{eq:sigma}) 
tells that T cells activate in the presence of antigen presenting cells and 
become inactive in the absence of antigens. However, biological noise may 
occasionally lead the system to deviate from the expected behaviour. 
For example, some T cells may fail to activate even in the presence of their 
conjugate antigen presenting cell, or, conversely, some cells may activate 
randomly, even in the absence of antigens (for example due to activation 
by self-antigens in an auto-immune response). Such stochastic effects are 
modelled by the random variable $z$: positive values of $z$ damp the immune
response while negative values 
enhance it.
The time step $\Delta$ will eventually be sent to zero to retrieve the continuous time dynamics.

We note that we could easily account for activation of T cells mediated by dendritic cells by 
adding a term $\sum_\mu \xi_i^\mu d_\mu$ in the argument of the step function, with $d_\mu$ denoting 
the density of active dendritic cells, however, we will show later that this extra term has only a small effect, hence, for simplicity, we will neglect it here. 

We note that equation (\ref{eq:sigma}) models reaction (\ref{eq:T_active}) in the presence of noise, at 
the {\it microscopic} level of individual T cells. 
Alternatively, one could model (\ref{eq:T_active}) at the 
population level, via reaction kinetics equations, similar to (\ref{eq:p_dyn}) and (\ref{eq:psi_dyn}), for the densities of active and inactive $T$ 
cells, valid under the assumptions of well-mixed system and 
negligible fluctuations due to discreteness of cells.
Noise could be included at population level, by introducing a reaction for 
spontaneous activation of $T$ cells of the type $T\to T^\star$ and one for 
spontaneous deactivation $T^\star \to T$, the rates of which would represent free parameters of the model. 
Our approach starts instead from stochastic equations for the microscopic cell states, which do not require 
the above assumptions and keep the number of free parameters to a minimum. 
Macroscopic cell densities as those involved in reaction kinetics approaches, can be obtained within our approach, 
as sums of microscopic variables, e.g. 
the density of active T cells binding APB $\mu$ can be obtained from $\sum_i \xi_i^\mu \sigma_i$, similarly the 
density of active helper T cells binding APB $\mu$ is obtained from $\sum_i \xi_i^\mu \sigma_i(1+\eta_i)/2$.

Finally, B clones expand (contract) 
when they receive excitatory (inhibitory) signals from active 
T cells and compete for survival, 
so that each B clone follows a logistic dynamics 
\bea
\frac{d b_\mu}{dt}&=&
b_\mu\left(\frac{\lambda_\mu^+}{c_\mu N^{1-\gamma}}\sum_{i=1}^N \xi_i^\mu \sigma_i \frac{1+\eta_i}{2}
- \frac{\lambda_\mu^-}{c_\mu N^{1-\gamma}}\sum_{i=1}^N \xi_i^\mu \sigma_i\frac{1-\eta_i}{2}
-\frac{\pi_\mu^+}{n}\psi_\mu +\frac{\pi_\mu^-}{n}p_\mu 
- \delta_\mu b_\mu
\!\right)
\label{eq:logistic_B}
\eea
Here, the first term represents clonal expansion, via 
reaction (\ref{eq:B_expand}),
triggered by active ($\sigma_i=1$) helper ($\eta_i=1$) 
T cells, specific for antigen $\mu$ ($\xi_i^\mu=1$). The second term 
accounts for clonal contraction, via reaction (\ref{eq:B_contract}),
triggered by active, specific suppressor ($\eta_i=-1$) T cells. 
The third 
and fourth 
term accounts for the binding 
and unbinding 
of B cells with antigens, 
respectively, 
via reaction (\ref{eq:APB}) and (\ref{eq:APB_remove}).  
Since antigen binding typically engages only one (or at most a few)
of the $n$ BCRs, the resulting APB is effectively still 
a B cell, with one spare receptor less. This leads to a 
decrease 
(increase) 
of the effective number of B 
cells by a fraction $1/n$ only, upon binding 
(unbinding) 
an antigen. 
The last term, a loss term proportional to the square of B cells population density, accounts for competition betweeen B cells.  
Different molecular mechanisms of clonal suppression have been described with no consensus on one universal mechanism \cite{Janeway}. 
For mathematical simplicity, we will assume 
that clonal suppression takes place at the same rate as clonal expansion, and set $\lambda_\mu^+=\lambda_\mu^-=\lambda_\mu$. Generalizations 
to the case $\lambda_\mu^+\neq \lambda_\mu^-$ are straightforward and we will comment on them later. This simplifies (\ref{eq:logistic_B}) to
\bea
\frac{d}{dt} b_\mu&=& 
b_\mu\left(\lambda_\mu m_\mu(\bsigma)- \delta_\mu  b_\mu
\right)
\label{eq:logistic_b}
\eea
where we have neglected $\order{(n^{-1})}$ terms, bearing in mind that $n=\order{(10^5)}$, and
defined the density of the net excitatory signal received by B clone $\mu$ 
from T cells as
\be
m_\mu(\bsigma)=\frac{1}{c_\mu N^{1-\gamma}}\sum_{i=1}^N \sigma_i \eta_i \xi_i^\mu
\label{eq:mmu}
\ee
with $\bsigma=(\sigma_1,\ldots,\sigma_N)$ representing the microstate (active or inactive) of all
T cells.

\subsection{Macroscopic dynamics}
The dynamics of the immune system model defined above entails 
the stochastic update (\ref{eq:sigma}) of the microstate $\bsigma$ of all 
T cells. 
However, cell concentrations 
are seen to depend on $\bsigma$ only through the variables 
$\bm(\bsigma)=(m_1(\bsigma), \ldots, 
m_P(\bsigma))$, via (\ref{eq:logistic_b}).
In this section, we will use 
non-equilibrium statistical mechanical techniques \cite{Coolen1, Coolen2,
Silvia_1} 
to derive, from the law of the microscopic system $\bsigma$, 
equations for the macroscopic variables $\bm(\bsigma)$. 
To this purpose, we write below the master equation for the 
probability density $p_t(\bsigma)$ to observe a
microstate $\bsigma$ at time $t$, and we will derive from this a dynamical 
equation for the 
macroscopic variables $\bm(\bsigma)$.

Denoting by ${\mathcal P}(z\leq x)=\int_{-\infty}^x dz\, p(z)$ the cumulative 
distribution function of the noise distribution $p(z)$, 
the likelihood to observe 
configuration $\sigma_i$ at time $t+\Delta$, for any symmetric distribution 
$p(z)=p(-z)$, is \cite{book_CKS}
$$
p_{t+\Delta}(\sigma_i)={\mathcal P}\left(z\leq (2\sigma_i-1) \sum_\mu \xi_i^\mu p_\mu(t)
\right)
$$ 
A natural choice for 
$p(z)$ would be a Gaussian distribution, with variance $\beta^{-1}$,
which leads to 
${\mathcal P}(z\leq x)=\half(1+ {\rm erf}(\beta x/\sqrt{2}))$. 
An alternative choice is the so-called Glauber distribution 
leading to  
\be
{\mathcal P}(z\leq x) = \frac{1}{2}
\left(1 + \tanh \frac{\beta x}{2}\right),
\label{eq:glauber}
\ee
which is qualitatively very similar to the cumulative distribution function 
for the Gaussian distribution and is easier to work with analytically. 
The parameter $\beta$ has to be interpreted as an inverse noise level: for 
$\beta\to 0$, the dynamics (\ref{eq:sigma}) is fully stochastic, whereas for $\beta \to \infty$, the 
dynamics is deterministic.
For the choice (\ref{eq:glauber}), the probability that 
helper T cell $i$ {\it changes},
in a single time step, its state $\sigma_i$ at time $t$ (to $1-\sigma_i$
at time $t+\Delta$) is
\be
W_t(\sigma_i)=
\half \left[1+(1-2\sigma_i)\tanh \frac{\beta}{2} \sum_\mu \xi_i^\mu 
p_\mu(t)\right]
\label{eq:Wt}
\ee
where we used $1-2\sigma_i=\pm 1$ and 
$\tanh(\pm x)=\pm \tanh x$.
Assuming that the update of T cells is sequential, i.e. 
at each time step one helper cell $i$, drawn at random, is updated 
with likelihood $W_t(\sigma_i)$, one obtains, 
for $\Delta=1/N$ and $N$ large, the following master equation for 
the probability density $p_t(\bsigma)$ to observe 
microstate $\bsigma$ at time $t$, 
\be
\frac{d}{dt} p_t(\bsigma)=\sum_i [p_t(F_i \bsigma)W_t(1-\sigma_i)
-p_t(\bsigma)W_t(\sigma_i)]
\label{eq:master}
\ee
where $F_i$ is a ``cell-flip''
operator that changes the configuration of T cell $i$ from 
$\sigma_i$ to $1-\sigma_i$ and has no effect on any other cell $j\neq i$.
From (\ref{eq:master}) one can derive equations of motion for 
expectations
$\bra \cdot \ket=\sum_\bsigma \cdot\, p_t(\bsigma)$. 
Multiplying (\ref{eq:master}) by $\sigma_j$ and summing over $\bsigma$, 
we obtain the rate of change 
\be
\frac{d}{dt}\bra \sigma_j\ket=\bra (1-2\sigma_j)W_t(\sigma_j)\ket
\label{eq:av_s}
\ee
of the average activity of T cell $j$, intuitively  
given by its variation 
$1-2\sigma_j$ upon a single cell flip $F_j$, times the rate $W_t(\sigma_j)$ 
at which the cell is flipped.  

Then, multiplying (\ref{eq:av_s}) times $\eta_j \xi_j^\mu$, summing over $j$,
dividing by $c_\mu N^{1-\gamma}$, and using a mean-field approximation 
(see \ref{app:KM}), we obtain the following equation of motion for 
the average signal strength 
$m_\mu(t)=\bra m_\mu(\bsigma(t))\ket$ on clone $\mu$:
\be
\frac{dm_\mu}{dt}=\frac{N^\gamma}{2c_\mu}\bra \xi^\mu \eta[1+\tanh 
\frac{\beta}{2} 
\sum_\nu \xi^\nu p_\nu ]\ket_{\eta,{\bxi}}-m_\mu
\label{eq:av_m}
\ee
In the above $\bra \dots \ket_{\eta, \bxi}$ denotes the average 
$\sum_{\eta,\bxi}\dots P(\eta,\bxi)$ over the distribution 
of regulatory patterns in the system
\be
P(\eta,\bxi) = 
\frac{1}{N}
\sum_{j=1}^N \delta_{\eta,\eta_j}\delta_{\bxi,\bxi_j}
\label{eq:joint}
\ee
where $\delta_{x,y}=1$ for $x=y$ and $\delta_{x,y}=0$ otherwise, 
and $\bxi_j=(\xi_j^1,\dots, 
\xi_j^P)$ encodes the regulatory interactions between B cells 
and T cell $j$.
Assuming that $P(\eta,\bxi)=P(\eta)P(\bxi)$ i.e. the ability 
of a helper cell $i\,$ to bind to clone $\mu$ does not depend on whether 
$\,i\,$ is a helper or regulator, we have
\be
\frac{dm_\mu}{dt}=\frac{\epsilon N^\gamma}{2c_\mu}\bra \xi^\mu [1+\tanh 
\frac{\beta}{2} 
\sum_\nu \xi^\nu p_\nu ]\ket_{\bxi}
-m_\mu.
\label{eq:many_ag}
\ee
Finally, averaging over $\xi^\mu$, using the independence of the $\xi^\mu$'s and $\bra \xi^\mu\ket =c_\mu/N^\gamma$, we get 
\be
\frac{dm_\mu}{dt}\!=\!\frac{\epsilon}{2}\!\left[1\!+\!\bra \tanh 
\frac{\beta}{2}\Big(p_\mu \!+\! 
\sum_{\nu\neq \mu}^P \xi^\nu p_\nu\!\Big)\ket_{\{\xi^\nu\}} \!\right]
-m_\mu,
\label{eq:m_many_ag}
\ee
where one has still to carry out the average over the $\{\xi^\nu\}$ other 
than $\xi^\mu$. The second term in the round brackets represents clonal 
interference, i.e. contributions to clone $\mu$ arising from 
different clones $\nu$'s, due to the ability of T cells 
to bind different B clones, as illustrated in Fig. \ref{fig:xi_mu}. 
Due to the specificity of the interactions between B and T cells, however,
the $\xi$'s are extremely diluted, 
i.e. the probability for each $\xi^\mu$ to be non-zero 
is $\order{(N^{-\gamma})}$. Since only non-zero $p_\nu$'s contribute 
to clonal interference, as long as the number of different 
antigenic threats in the host is finite, 
the sum over $\nu$ consists of a finite number of terms, each 
$\order{(N^{-\gamma})}$, and 
vanishes in the thermodynamic limit. Clonal interference becomes instead 
$\order{(1)}$ when the number of different 
antigens in the host is $\order{(N^\gamma)}$, i.e. of 
the same order as the number of B clones. 
In the remainder of the paper, we will focus on the immune response when a 
single antigen is present in the host, so cross-reactivity effects 
between B and T cells can be mostly neglected, however we will see  
in Sec. \ref{sec:cross-r}, that they may cumulate with cross-reactivity 
effects betweeen B cells and antigens, when present.

In \ref{app:KM} we show that for $\gamma<1$ fluctuations of $m_\mu(\bsigma)$ 
about its thermodynamic average $m_\mu$ vanish for large $N$. In this regime, the mean-field approximation becomes exact and 
we are allowed to replace $m_\mu(\bsigma)$ in (\ref{eq:logistic_b}) with $m_\mu$ 
\bea
\frac{db_\mu}{dt} &=&
b_\mu\left(\lambda_\mu m_\mu- \delta_\mu  b_\mu
\right)
\label{eq:b_dyn}
\eea
which enables us to express the dynamical evolution 
of our model in terms of a closed 
set of first order differential 
equations, namely (\ref{eq:p_dyn}), (\ref{eq:psi_dyn}), 
(\ref{eq:m_many_ag}) and (\ref{eq:b_dyn}).

We conclude this section by noting that B clonal dynamics (\ref{eq:b_dyn}) has one fixed point at $b_\mu=\lambda_\mu m_\mu/\delta_\mu$,
which is stable for $m_\mu>0$, and one fixed point at $b_\mu=0$ which is 
stable for $m_\mu\leq 0$, suggesting that the immune system must keep a 
basal activity ($m_\mu>0$), even in the absence of antigens, to sustain 
cell numbers. In the absence of antigens, 
non-zero values of (\ref{eq:mmu}) are achieved via stochastic 
fluctuations in T cell activation. 
Such activity, whereby T cells keep sending survival signals to 
B clones in the absence of antigens, is experimentally observed and is
believed to be one of the mechanisms to accomplish a homeostatic 
control of cell numbers \cite{homeostasis}.
In addition, lymphocyte homeostasis requires that clonal expansion 
during the immune 
response is matched by a comparable decrease 
in lymphocyte numbers once the antigen is removed. Its mechanics is not fully 
understood, and is believed to require a sophisticated regulation of 
the rates of cellular proliferation and programmed cell death. 
In this model, however, clonal contraction upon antigen removal, 
will emerge as a simple consequence of 
stochasticity in T cell activation, as we will see in the next section. 
As the antigen concentration is decreased, stochasticity effects become more 
important and eventually dominate the dynamics (\ref{eq:sigma}), thus 
restoring the activity (\ref{eq:mmu}) to its basal value, which
determines the equilibrium sizes of B clones.
%

\subsection{Response to a single antigen}
In this section, we study the activation of the immune system when a single 
antigen of type $\nu$ 
is present, i.e. $\psi_\nu \neq 0$ and $\psi_\mu=0~\forall~\mu \neq \nu$.
We assume that 
the system is initially at equilibrium in the absence of 
antigens, when antigen $\nu$ is introduced. 

The equilibrium state in the absence of antigen is easily found by noting that $\psi_\mu=0$ is a fixed 
point of the $\mu$-antigen dynamics (\ref{eq:psi_dyn}), and equation (\ref{eq:p_dyn}) implies that at stationarity 
the density of B cells presenting antigen $\mu$ is
\be
p_\mu=a_\mu b_\mu \psi_\mu
\label{eq:p_stat}
\ee
where $a_\mu=\pi^+_\mu/\pi_\mu^-$ can be thought of as a measure of the affinity between 
B clone $\mu$ and antigen $\mu$.\footnote{The affinity can be 
experimentally measured 
as the ratio between the equilibrium concentration of APB and the product of 
the 
equilibrium concentrations of antigen and conjugate B cells. }
Hence, at equilibrium, in the absence of antigen, 
one has $\psi_\mu=0$, $p_\mu=0~\forall~\mu$ and, from (\ref{eq:m_many_ag}), $m_\mu=\epsilon/2~\forall~\mu$.  
Then (\ref{eq:b_dyn}) yields $b_\mu=\kappa_\mu \epsilon/2$, with 
$\kappa_\mu=\lambda_\mu/\delta_\mu$, for $\epsilon>0$,
and $b_\mu=0$ for $\epsilon<0$. 

\noindent
Upon introducing antigen $\nu$, equation (\ref{eq:m_many_ag}) gives, 
for any $\mu\neq \nu$   
\bea
\frac{dm_\mu}{dt}&=&\frac{\epsilon}{2}\left[
1+\bra \tanh\frac{\beta}{2}\xi^\nu p_\nu\ket_{\xi^\nu}
\right]-m_\mu
\nonumber\\
&=&\frac{\epsilon}{2}-m_\mu +\order{(N^{-\gamma})},
\label{eq:m_mu}
\eea
showing that, for large $N$, 
$m_\mu=\epsilon/2$ is a stable fixed point for 
all clones $\mu\neq \nu$ not responding to antigen $\nu$, which thus 
permane in the state 
\be
(\psi_\mu, p_\mu, m_\mu, b_\mu)=\left(0,0,\frac{\epsilon}{2},{\rm max}\left\{0,\frac{\kappa_\mu \epsilon}{2}\right\}\right).
\nonumber\ee
In contrast, 
for the $\nu$-clone, conjugate to the antigen introduced in the host,
(\ref{eq:m_many_ag}) gives
\bea
\frac{dm_\nu}{dt}&=&\frac{\epsilon}{2}\!\left[1\!+\!\tanh 
\frac{\beta}{2}p_\nu  \!\right]
-m_\nu
\label{eq:m_one_ag}
\eea
which has to be solved together with 
\bea
\frac{d p_\nu}{dt}&=&\pi_\nu^+ \psi_\nu b_\nu-\pi_\nu^- p_\nu
\label{eq:dyn_sys1}
\\
\frac{d \psi_\nu}{dt}&=&\psi_\nu(r_\nu-\pi_\nu^+  b_\nu)
\label{eq:dyn_sys2}
\\
\frac{db_\nu}{dt}&=&\delta_\nu b_\nu(\kappa_\nu m_\nu-b_\nu)
\label{eq:dyn_sys3}
\eea
leading to a fourth order dynamical system, with Jacobian 
\be
J=\left(
\begin{array}{cccc}
-1 & \frac{\beta\epsilon}{4}\left(1-\tanh^2\frac{\beta p_\nu}{2}\right)
& 0 & 0 
\\
0 & -\pi_\nu^- & \pi_\nu^+ b_\nu & \pi_\nu^+ \psi_\nu
\\
0 & 0 & r_\nu-\pi_\nu^+  b_\nu & -\pi_\nu^+ \psi_\nu
\\
\delta_\nu b_\nu \kappa_\nu & 0 & 0 & \delta_\nu (\kappa_\nu m_\nu -2b_\nu)
\end{array}
\right)
\ee
The $\nu$-clone will return to its resting state, meaning that the antigen will be cleared, 
if the state $(\psi_\nu, p_\nu, m_\nu, b_\nu)=\left(0,0,\frac{\epsilon}{2},{\rm max}\left\{0,\frac{\kappa_\nu \epsilon}{2}\right\}\right)$
is a stable fixed point of the $\nu$-clone dynamics (\ref{eq:m_one_ag}, \ref{eq:dyn_sys1}, 
\ref{eq:dyn_sys2}, \ref{eq:dyn_sys3}).
This requires the eigenvalues of the Jacobian, evaluated at the resting state, to be negative. 
For $\epsilon>0$, the resting state is 
$\left(0,0,\frac{\epsilon}{2},\frac{\kappa_\nu \epsilon}{2}\right)$ 
and the eigenvalues 
\bea
\lambda_1&=&-1
\\
\lambda_2&=&-\pi_\nu^-
\\
\lambda_3&=&r_\nu-\pi_\nu^+ \kappa_\nu\frac{\epsilon}{2}
\\
\lambda_4&=&-\frac{\lambda_\nu \epsilon}{2}
\eea 
are all negative for 
$\epsilon>2r_\nu/(\pi_\nu^+ \kappa_\nu)$, meaning that the antigen will be cleared 
when the 
affinity $\pi_\nu^+$ of the conjugate B cells is greater than a value
that increases as the pathogen replication rate 
increases and $\epsilon$ decreases.
For $\epsilon<0$, the Jacobian must be evaluated at the state 
$\left(0,0,\frac{\epsilon}{2},0\right)$ and the corresponding eigenvalues
\bea
\lambda_1&=&-1<0
\\
\lambda_2&=&-\pi_\nu^-<0
\\
\lambda_3&=&r_\nu>0
\\
\lambda_4&=&-\frac{\lambda_\nu |\epsilon|}{2}<0
\eea 
show an instability in the antigen dynamics. 
It is easy to see from (\ref{eq:m_one_ag}) and (\ref{eq:dyn_sys3}), 
that in this regime $m_\nu$ will converge to 
negative values, as $\tanh(x)\geq 0~\forall~x\geq 0$, and $b_\nu$ will vanish in the long-time limit,  
leading to an immuno-suppressed host, where the antigen grows indefinitely according to
$$
\frac{d}{dt}\psi_\nu=r_\nu\psi_\nu.
$$
Finally, for $0<\epsilon<2r_\nu/(\kappa_\nu \pi_\nu^+)$ 
the fixed point $(\psi_\nu, p_\nu, m_\nu, b_\nu)=\left(0,0,\frac{\epsilon}{2},
\frac{\kappa_\nu \epsilon}{2}\right)$ is unstable and different types 
of dynamics may 
arise, depending on the range of the kinetic parameters. 
Equations (\ref{eq:m_one_ag}), (\ref{eq:dyn_sys1}) and (\ref{eq:dyn_sys3}) 
show that, 
in this regime, $m_\nu$ will approach positive values in the long-time limit 
and $b_\nu$ and $p_\nu$ will thus evolve towards $\kappa_\nu m_\nu$ and $a_\nu \kappa_\nu m_\nu \psi_\nu$, respectively.
Hence, in the long-time limit, the dynamics of clone $\nu$ can be described in terms of the second order dynamical system 
\bea
\frac{d}{dt}m_\nu&=&\frac{\epsilon}{2}\left[
1+\tanh\frac{\beta}{2}a_\nu \kappa_\nu m_\nu\psi_\nu
\right]-m_\nu
\label{eq:m_stat}
\\
\frac{d}{dt} \psi_\nu&=&\psi_\nu\left(r_\nu-\kappa_\nu \pi_\nu^+ m_\nu\right).
\label{eq:psi_m}
\eea
This has a fixed point at 
\bea
m_\nu^\star&=&\frac{r_\nu}{\kappa_\nu \pi_\nu^+}
\nonumber\\
\psi_\nu^\star&=&\frac{2 \pi_\nu^+}{a_\nu r_\nu \beta}\atanh\left(
\frac{2r_\nu}{\epsilon \kappa_\nu \pi_\nu^+} -1
\right)
\label{eq:psi_T}
\eea
whose stability is determined from the eigenvalues 
\be
\lambda_{1,2}=\frac{A_\nu\psi_\nu^\star-1}{2}\pm \sqrt{\left(\frac{A_\nu\psi_\nu^\star-1}{2}\right)^2-r_\nu A_\nu 
\psi_\nu^\star}
\ee
where
$$
A_\nu=\frac{\beta a_\nu r_\nu}{\pi_\nu^+}\left(1-\frac{r_\nu}{\epsilon \kappa_\nu \pi_\nu^+}
\right).
$$
We have
\be
A_\nu \psi_\nu^\star=\atanh\left(
\frac{2 r_\nu}{\epsilon \kappa_\nu \pi_\nu^+}-1\right)\left[
1-\left(
\frac{2 r_\nu}{\epsilon \kappa_\nu \pi_\nu^+}-1\right)
\right]<1
\ee
as $x\,\atanh x -\atanh x +1>0 ~\forall~x$, 
hence both eigenvalues $\lambda_{1,2}$ have negative real part for $A_\nu>0$ i.e. for
$\epsilon>r_\nu/(\kappa_\nu \pi_\nu^+)$.
Therefore, in the regime $r_\nu/(\pi_\nu^+\kappa_\nu)<\epsilon<2r_\nu/(\pi_\nu^+\kappa_\nu)$, 
the fixed point $(m_\nu^\star,\psi_\nu^\star)$ is stable and the 
antigen will approach the non-zero value (\ref{eq:psi_T}), which is higher, the higher the noise level $\beta^{-1}$,
while $m_\nu$, $b_\nu$ and $p_\nu$ will approach the $\beta$-independent values 
$r_\nu/(\kappa_\nu \pi_\nu^+)$, $r_\nu/ \pi_\nu^+$ and $a_\nu r_\nu \psi_\nu^\star/\pi_\nu^+$, respectively.
Conversely, for $0<\epsilon<r_\nu/(\kappa_\nu \pi_\nu^+)$ the fixed point $(m_\nu^\star,\psi_\nu^\star)$ is unstable.  
It is easy to see from (\ref{eq:m_stat}), that for long times $0 \leq m_\nu \leq \epsilon$, as
$-1\leq \tanh x \leq 1~\forall ~x$, then (\ref{eq:psi_m}) 
implies that when $m_\nu$ reaches its maximum value $\epsilon$, the viral concentration will keep 
growing in the regime $\epsilon<r_\nu/(\pi_\nu^+\kappa_\nu) $.
%
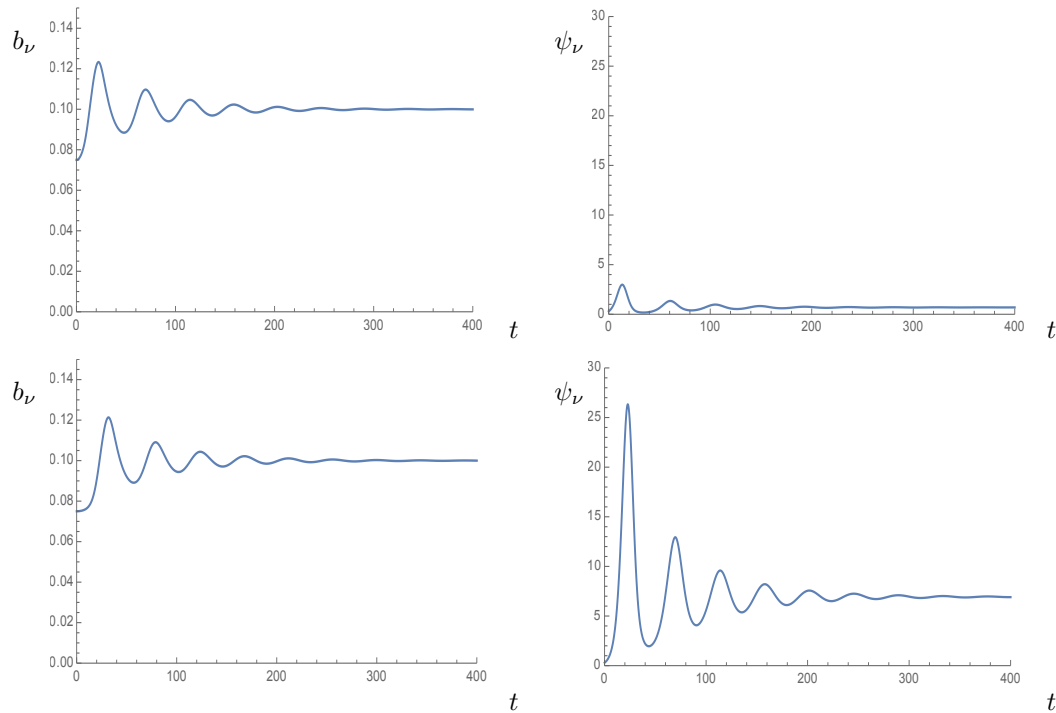
\begin{figure}
\hspace*{2cm}
\setlength{\unitlength}{0.55mm}
\begin{picture}(400,170)
\put(0,155){$b_\nu$}
\put(131,155){$\psi_\nu$}
\put(120,85){$t$}
\put(250,85){$t$}
\put(0,70){$b_\nu$}
\put(131,70){$\psi_\nu$}
\put(120,-5){$t$}
\put(250,-5){$t$}
\put(9,85){\includegraphics[width=105\unitlength,height=80\unitlength]{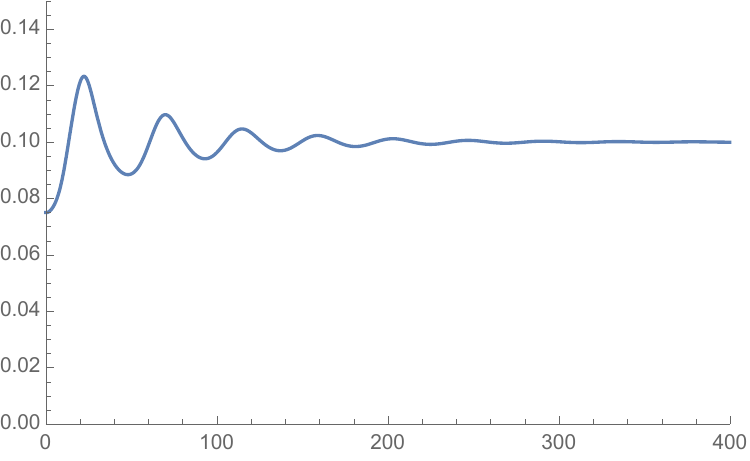}}
\put(140,85){\includegraphics[width=105\unitlength,height=80\unitlength]{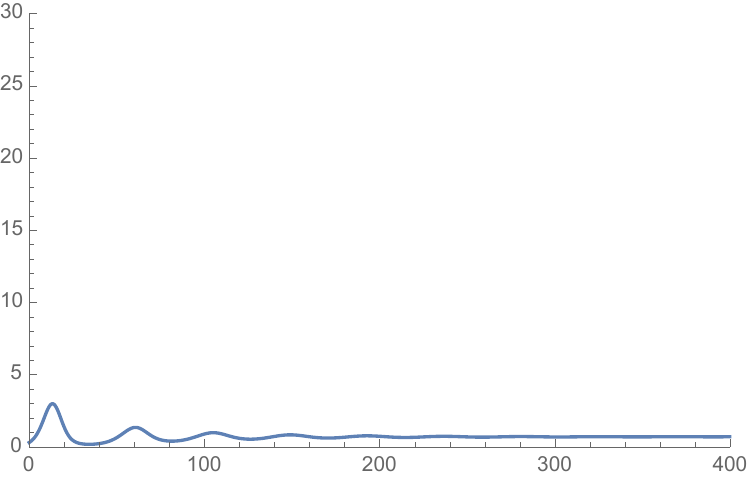}}
\put(9,0){\includegraphics[width=106\unitlength,height=80\unitlength]{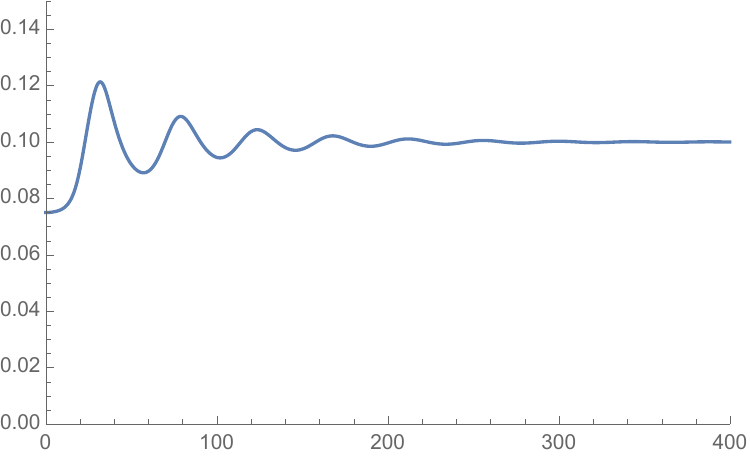}}
\put(139,0){\includegraphics[width=105\unitlength,height=80\unitlength]{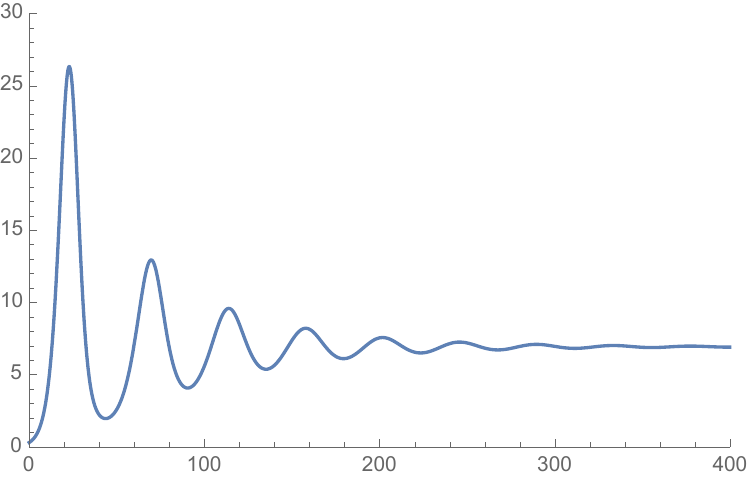}}
\end{picture}
\caption{Time evolution of B clonal density $b_\nu$ (left) and antigen 
concentration $\psi_\nu$ (right) for $r_\nu\!=\!\lambda_\nu\!=\!\delta_\nu\!=\!\pi_\nu^-\!=\!1$, $\epsilon\!=\!0.15$ and $\pi_\nu^+\!=\!10$.
Top panels: $\beta\!=\!1$. Bottom panels: $\beta\!=\!0.1$;
Initial conditions were chosen as $\psi_\nu(0)\!=\!0.3$, 
$b_\nu(0)\!=\!\kappa_\nu m_\nu(0)$, $m_\nu(0)\!=\!\epsilon/2$ and $p_\nu(0)=0$. 
As expected, the antigen concentration converges to a value that increases with the noise  
$\beta^{-1}$, while the 
B clonal density evolves to the $\beta$-independent value $b_\nu\!=\!r_\nu/\pi_\nu^+ $. 
} 
\label{fig:regime_ii}
\end{figure}
%
\begin{figure}
\setlength{\unitlength}{0.55mm}
\hspace*{2cm}
\begin{picture}(200,90)
\put(0,70){$b_\nu$}
\put(131,70){$\psi_\nu$}
\put(100,0){$t$}
\put(240,0){$t$}
\put(10,5){\includegraphics[width=105\unitlength,height=80\unitlength]{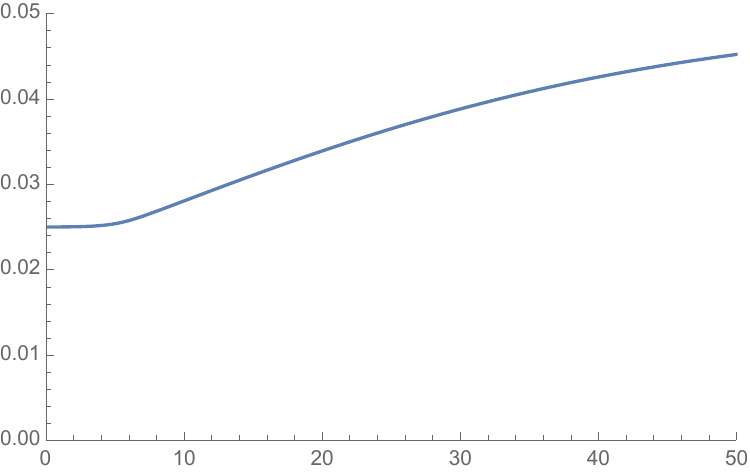}}
\put(140,5){\includegraphics[width=105\unitlength,height=80\unitlength]{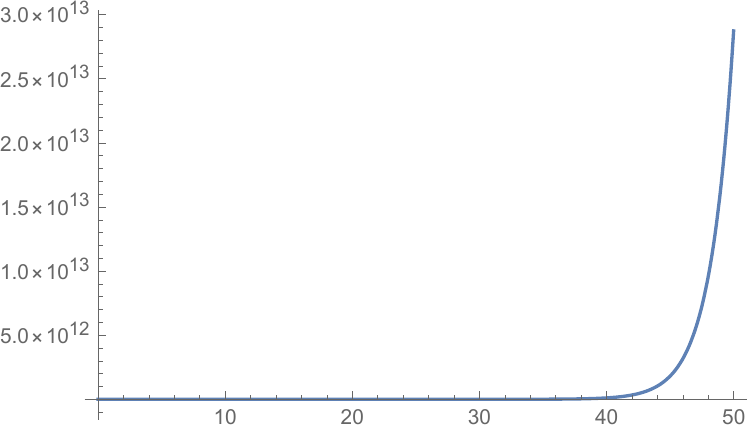}}
\end{picture}
\caption{Time evolution of B clonal density $b_\nu$ (left) and antigen 
concentration $\psi_\nu$ (right) for $r_\nu=\lambda_\nu=\delta_\nu=\pi_\nu^-=\beta=1$, $\epsilon=0.05$. 
Initial conditions were chosen as in Figure \ref{fig:regime_ii}.  
As expected, the B clonal density evolves to $b_\nu=\kappa_\nu m_\nu $ with $m_\nu$ attaining its maximum value $\epsilon$. 
} 
\label{fig:regime_iii}
\end{figure}
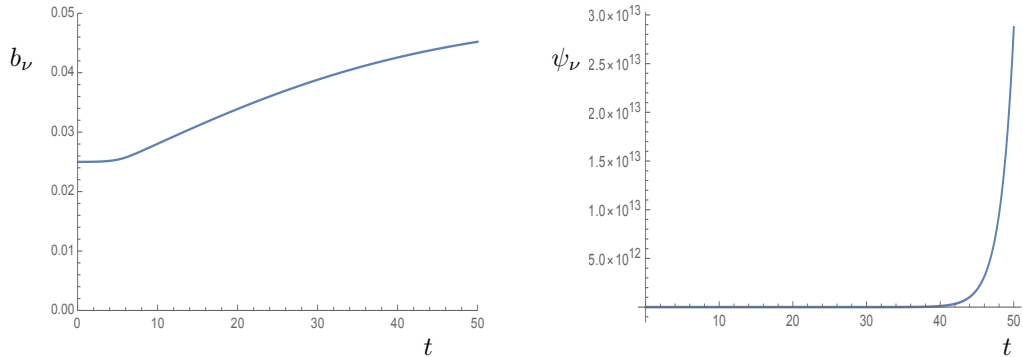
%
%
%
Numerical solutions of the full dynamical system (\ref{eq:m_one_ag}, \ref{eq:dyn_sys1}, \ref{eq:dyn_sys2}, \ref{eq:dyn_sys3})
are plotted in Figure \ref{fig:regime_ii} for $r_\nu/(\kappa_\nu \pi_\nu^+)<\epsilon<2r_\nu/(\kappa_\nu \pi_\nu^+)$
and in Figure \ref{fig:regime_iii} for $0<\epsilon<r_\nu/(\kappa_\nu \pi_\nu^+)$, and confirm these 
predictions, which can also be verified by evaluating numerically 
the eigenvalues of the full dynamical system, in these regimes.

In conclusion, four different regimes are identified: (i) $\epsilon>2r_\nu/(\kappa_\nu \pi_\nu^+)$ 
where the antigen is cleared and the responding B clone returns to its resting size;  
(ii) $r_\nu/(\kappa_\nu \pi_\nu^+)<\epsilon<2r_\nu/(\kappa_\nu \pi_\nu^+)$ where the viral concentration evolves to a non-zero value, 
which, interestingly, depends on the noise level. In this regime the antigen permanes indefinitely in the host, 
although its growth is limited by the action of the immune system. The responding B clone remains in an expanded state $m_\nu^\star>\epsilon/2$ and fails to contract and return to its resting size, even at large times; 
(iii) $0<\epsilon<r_\nu/(\kappa_\nu \pi_\nu^+)$ where the immune system responds to the antigen by expanding the 
conjugate B clone, but the response is too weak and the antigen proliferates indefinitely in the host; 
(iv) $\epsilon<0$ where the immune system is irresponsive and B cells decrease over time, while the antigen proliferates indefinitely.
Finally we note that including the contribution from dendritic cells to T cells 
activation, would simply add a term in the argument of the hyperbolic 
tangent in (\ref{eq:m_one_ag}).
Since $-1\leq \tanh x \leq 1 ~\forall~x$, 
the inclusion of dendritic cells would not alter the different phases 
and would only have a small 
quantitative effect on the transient response.

\subsubsection{The role of affinity maturation}
We have so far regarded the binding rate between antigen and B cells $\pi_\nu^+$ as a constant, however, 
as B cells undergo clonal expansion, they increase their affinity with the antigen by several orders of magnitude \cite{Affinity}, via affinity
maturation, so $\pi_\nu^+$ is an increasing function of time \cite{B_GC}. Several studies have modelled in detail B cell maturation in the germinal centre, 
over the last few decades \cite{referee1, referee2, referee3}. Here, we simply assume that the affinity is an increasing function of time 
that saturates at large time, and choose $\pi_\nu^+(t)$ as the sigmoid function
\be
\pi^+_\nu(t)=\pi_M [\pi_0+\tanh(v(t-t^\star))]
\label{eq:pip}
\ee
where $v, t^\star, \pi_M$ and $\pi_0$ are parameters that 
control, respectively, how fast, early and large the affinity grows and its initial value. 
Experimental findings suggest that high affinity B cells bind their target antigen within a few minutes \cite{Factors}, 
hence we estimate $\pi_\nu^+(\infty)=\pi_M(\pi_0+1)$ to be $\order{(10^2)}/{\rm day}$. We choose the remaining parameters in such a way that the ratio 
$\pi_\nu^+(\infty)/\pi_\nu^+(0)=(\pi_0+1)/(\pi_0-\tanh v t^\star)$ between high and low affinity 
is within the estimated physiological range $\order{(10^5)}$ 
\cite{physiological}. 
It is also suggested that birth and death rates of B cells in the germinal centre are of the same order of magnitute, 
yielding $\kappa_\nu=\order{(1)}$ \cite{kappa}. 
This implies that regimes (ii) and (iii) take place on a narrow 
range of values of $\epsilon\in (0,2r_\nu/(\kappa_\nu \pi_\nu^+(\infty)))$:
above this range, the antigen is cleared, while below it, 
the immune system is irresponsive and the antigen replicates indefinitely in the host, no matter how fast or large $\pi_\nu^+$ grows. 
This results in a rather abrupt transition of the immune system from a responsive to a suppressed state, as 
$\epsilon$ approaches zero from above, in line with experimental findings.

Numerical solution of (\ref{eq:m_one_ag}), (\ref{eq:dyn_sys1}), 
(\ref{eq:dyn_sys2}) and 
(\ref{eq:dyn_sys3}), in the regime $\epsilon\!>\! 2r_\nu/(\kappa_\nu \pi_\nu^+(\infty))$
are shown in Figs. \ref{fig:beta}, \ref{fig:a}, \ref{fig:kappa} and \ref{fig:r_pim}. Plots of
$b_\nu$ and $\psi_\nu$ versus time, are shown for
different choices of the inverse noise level $\beta$ (Fig. \ref{fig:beta}), time-dependence of $\pi_\nu^+$
(Fig. \ref{fig:a}), kinetic parameters $\lambda_\nu$, $\delta_\nu$ (Fig. \ref{fig:kappa}), 
and $\pi_\nu^-$, $r_\nu$ (Fig. \ref{fig:r_pim}).
%
\begin{figure}[t!]
\hspace*{2cm}
\setlength{\unitlength}{0.55mm}
\begin{picture}(200,170)
\put(0,155){$b_\nu$}
\put(131,155){$\psi_\nu$}
\put(120,85){$t$}
\put(250,85){$t$}
\put(0,70){$b_\nu$}
\put(131,70){$\psi_\nu$}
\put(120,-5){$t$}
\put(250,-5){$t$}
\put(9,85){\includegraphics[width=105\unitlength,height=80\unitlength]{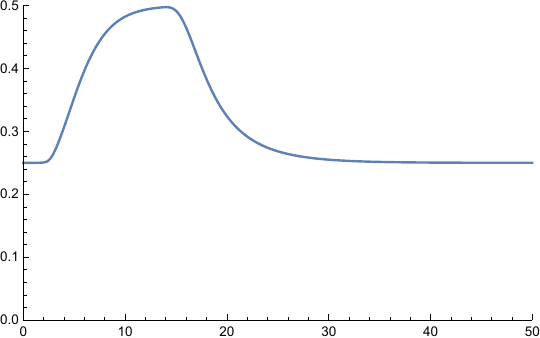}}
\put(140,85){\includegraphics[width=105\unitlength,height=80\unitlength]{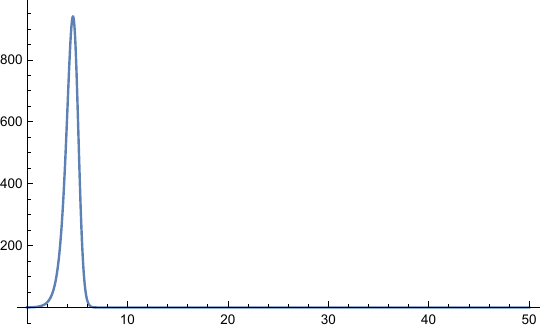}}
\put(9,0){\includegraphics[width=106\unitlength,height=80\unitlength]{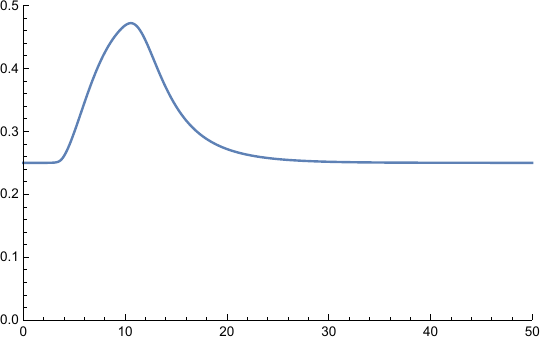}}
\put(139,0){\includegraphics[width=105\unitlength,height=80\unitlength]{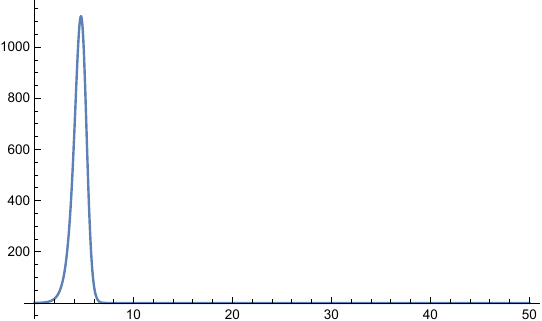}}
\end{picture}
\caption{Time evolution of B clonal density $b_\nu$ (left) and antigen 
concentration $\psi_\nu$ (right) for $r_\nu=2$, $\epsilon=0.5$, 
$\lambda_\nu=\delta_\nu=\pi_\nu^-=1$ and $\pi_\nu^+=10\times[\tanh(10)+\tanh(t-5)]$.
Top panels: $\beta=10$. Bottom panels: $\beta=0.1$;
Initial conditions were chosen as in Figure \ref{fig:regime_ii}.
For higher noise levels $\beta^{-1}$, the system mounts a weaker immune 
response, resulting in lower B clonal densities and higher 
antigen concentrations (note the different scales in the plots on the right),
however the antigen is still removed within the same timescale. 
} 
\label{fig:beta}
\end{figure}
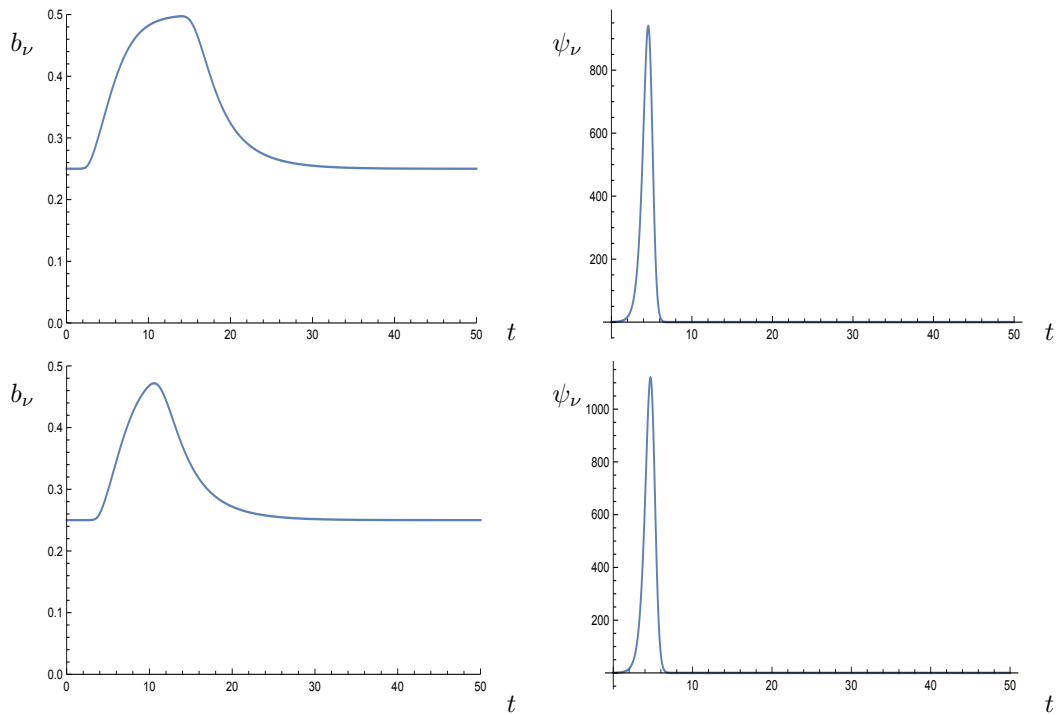
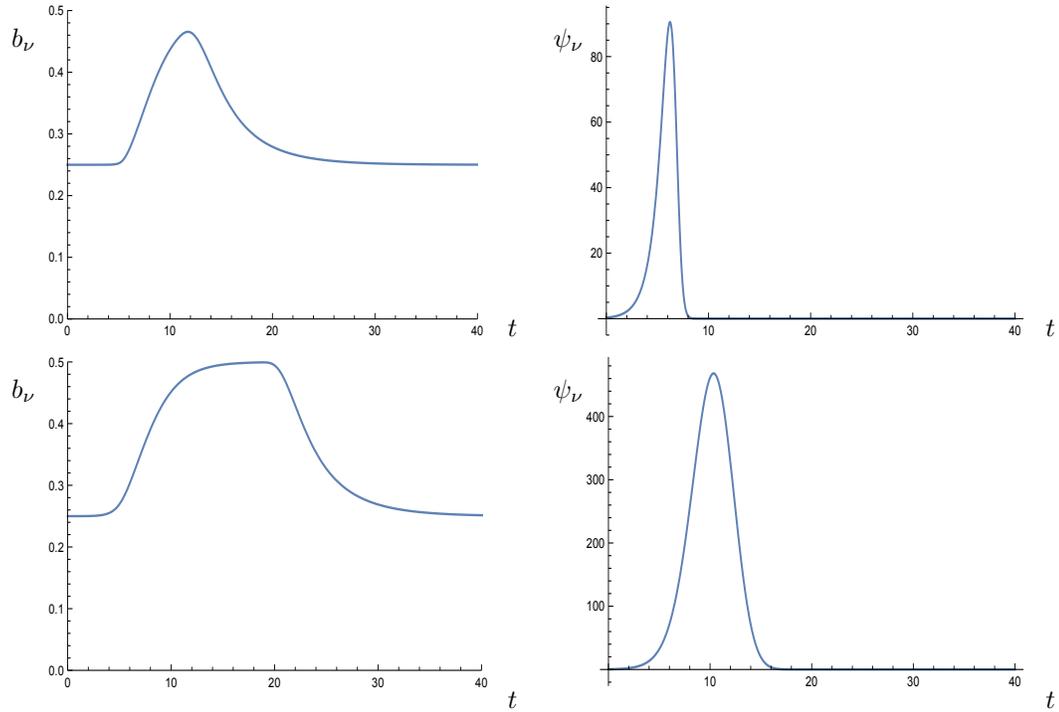
\begin{figure}[h!]
\hspace*{2cm}
\setlength{\unitlength}{0.55mm}
\begin{picture}(200,170)
\put(0,155){$b_\nu$}
\put(131,155){$\psi_\nu$}
\put(120,85){$t$}
\put(250,85){$t$}
\put(0,70){$b_\nu$}
\put(131,70){$\psi_\nu$}
\put(120,-5){$t$}
\put(250,-5){$t$}
\put(9,85){\includegraphics[width=105\unitlength,height=80\unitlength]{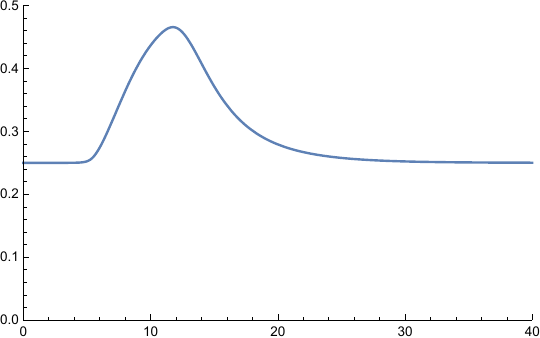}}
\put(140,85){\includegraphics[width=105\unitlength,height=80\unitlength]{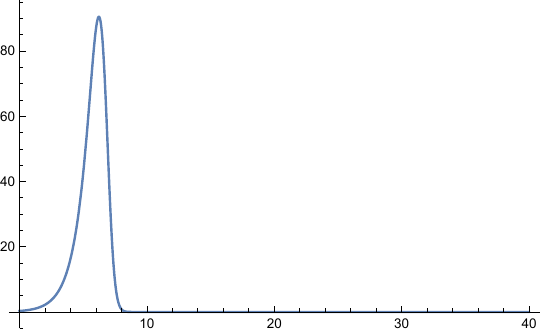}}
\put(9,0){\includegraphics[width=106\unitlength,height=80\unitlength]{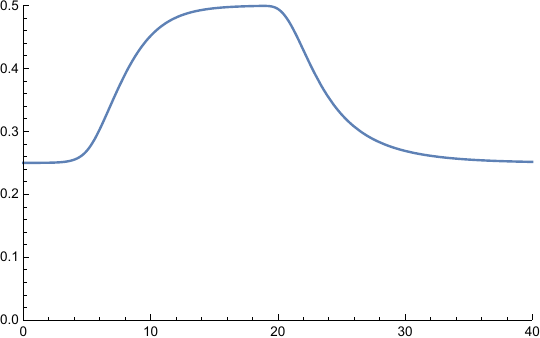}}
\put(139,0){\includegraphics[width=106\unitlength,height=80\unitlength]{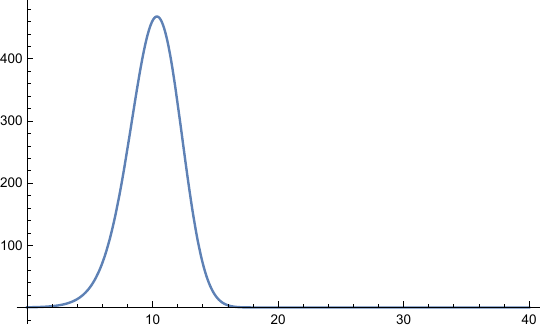}}
\end{picture}
\caption{Time evolution of B clonal density $b_\nu$ (left) and antigen 
concentration $\psi_\nu$ (right) for $r_\nu=1$, 
$\epsilon=0.5$, $\lambda_\nu=\delta_\nu=\pi_\nu^-=1$ and $\beta=1$.
Top panels: $\pi_\nu^+=10\times[\tanh(10)+\tanh(t-7)]$. Bottom panels: 
$\pi_\nu^+=10\times[\tanh(2.01)+\tanh[(t-20)/10]]$.
Initial conditions were chosen as in Figure \ref{fig:regime_ii}.
The time-dependence of the
affinity maturation $\pi_\nu^+$ affects both the intensity 
of viral concentration and the timescale on 
which viral removal is accomplished. Note that the plots on the right have very different scales.
} 
\label{fig:a}
\end{figure}
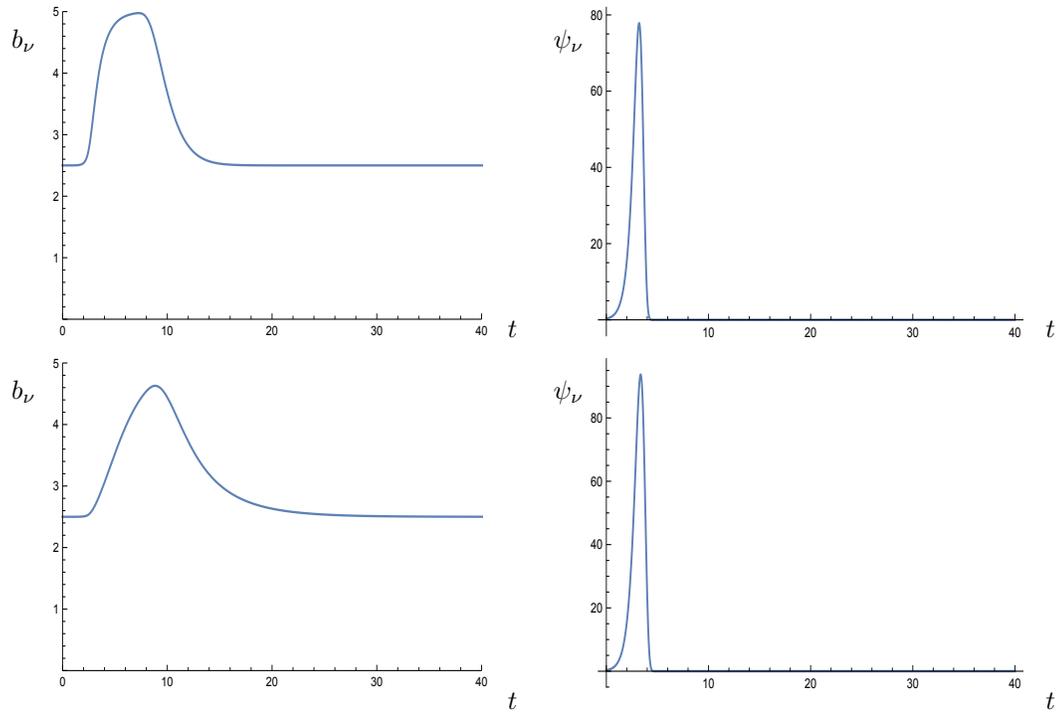
\begin{figure}[t!]
\hspace*{2cm}
\setlength{\unitlength}{0.55mm}
\begin{picture}(200,170)
\put(0,155){$b_\nu$}
\put(131,155){$\psi_\nu$}
\put(120,85){$t$}
\put(250,85){$t$}
\put(0,70){$b_\nu$}
\put(131,70){$\psi_\nu$}
\put(120,-5){$t$}
\put(250,-5){$t$}
\put(10,85){\includegraphics[width=105\unitlength,height=80\unitlength]{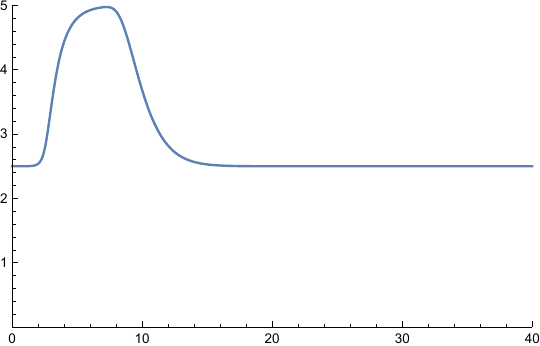}}
\put(140,85){\includegraphics[width=105\unitlength,height=80\unitlength]{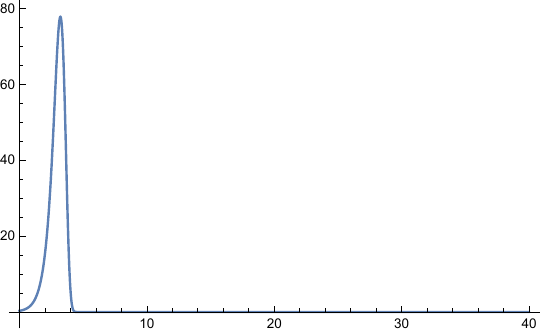}}
\put(10,0){\includegraphics[width=105\unitlength,height=80\unitlength]{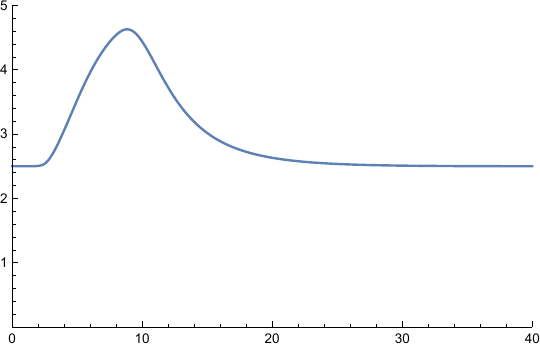}}
\put(140,0){\includegraphics[width=105\unitlength,height=80\unitlength]{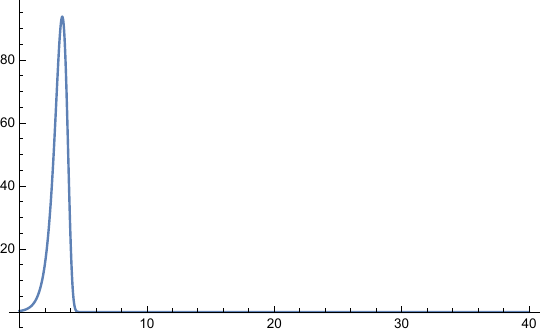}}
\end{picture}
\caption{Time evolution of B clonal density $b_\nu$ (left) and antigen concentration $\psi_\nu$ (right)
for $\beta=1$, $\pi_\nu^+=10\times[\tanh(10)+\tanh(t-5)]$, $\pi_\nu^-=1$, $r_\nu=2$ and $\epsilon=0.5$.
Top panels: $\lambda_\nu =10$ and $\delta_\nu=1$.
Bottom panels: $\lambda_\nu=1$, $\delta_\nu=0.1$.
Initial conditions were chosen as in Figure \ref{fig:regime_ii}. When compared to figure \ref{fig:beta}, these plots show that increasing the replication 
rate $\lambda_\nu$ of B cells or decreasing their competition $\delta_\nu$ leads to an increase in the B cell densities, but the antigen is still removed within similar 
timescales.
} 
\label{fig:kappa}
\end{figure}
\begin{figure}[h!]
\hspace*{2cm}
\setlength{\unitlength}{0.55mm}
\begin{picture}(200,170)
\put(0,155){$b_\nu$}
\put(131,155){$\psi_\nu$}
\put(120,85){$t$}
\put(250,85){$t$}
\put(0,70){$b_\nu$}
\put(131,70){$\psi_\nu$}
\put(120,-5){$t$}
\put(250,-5){$t$}
\put(10,85){\includegraphics[width=105\unitlength,height=80\unitlength]{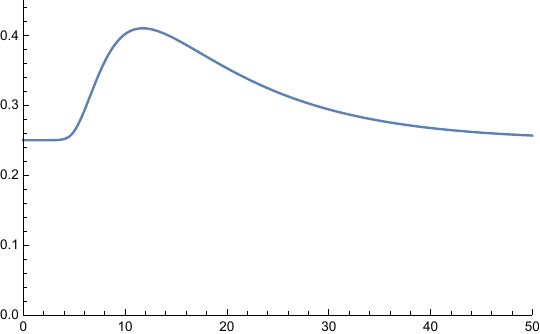}}
\put(146,85){\includegraphics[width=105\unitlength,height=80\unitlength]{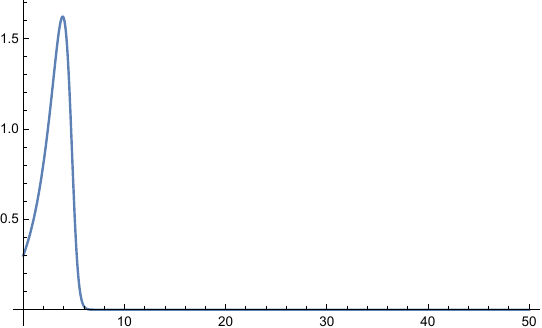}}
\put(11,0){\includegraphics[width=106\unitlength,height=80\unitlength]{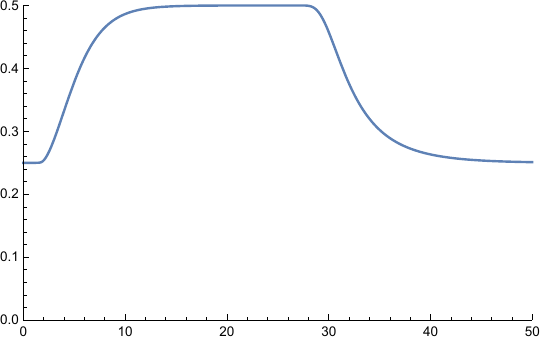}}
\put(140,0){\includegraphics[width=105\unitlength,height=80\unitlength]{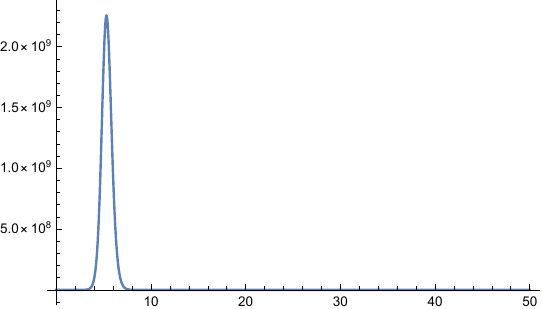}}
\end{picture}
\caption{Time evolution of B clonal density $b_\nu$ (left) and antigen concentration $\psi_\nu$ (right)
for $\beta=1$, $\pi_\nu^+=10\times[\tanh(10)+\tanh(t-5)]$, 
$\lambda_\nu =\delta_\nu=1$ and $\epsilon=0.5$.
Top panels: $r=0.5$ and $\pi_\nu^-=0.1$. 
Bottom panels: $r=5$, $\pi_\nu^-=1$. Initial conditions were chosen as in Figure \ref{fig:regime_ii}. A slower decay of B cells 
after the infection peak can be appreciated when decreasing $\pi_\nu^-$, which can potentially sustain long-term memory.
} 
\label{fig:r_pim}
\end{figure}
%
%
Plots show an immune response that follows closely the behaviour of the 
antigen, i.e. it expands B clones while the antigen 
concentration is increasing and 
contracts them when the antigen concentration is 
decreasing, 
so that both B clone 
and antigen concentrations are unimodal 
functions of time, meaning that the system is able to accomplish homeostasis. 
Interestingly, the time-dependent $b_\nu$ concentration shows an initial 
plateau, which is consistent with the experimentally observed lag-time 
between an infection and a detectable immune response \cite{Janeway}. 
The stochastic noise $\beta^{-1}$ 
has the effect of mildly
reducing the height of the peak in B clones concentration, thus 
increasing the peak in viral concentration, however the system will 
be able to remove the antigen even at high noise levels 
(see Fig. \ref{fig:beta}). 
The time-dependence of $\pi^+_\nu$,
affects both the 
location and the height of the peaks, consistently with 
the intuition that the faster the affinity grows, 
the earlier and the smaller the peak in viral concentration.
Fig. \ref{fig:a} shows that affinities increasing faster with time
($t^\star\!=\!7$, $v\!=\!1$) outperform
those that increase slower ($t^\star\!=\!20$, $v\!=\!0.1$). 
The kinetic parameters $\lambda_\nu$ and 
$\delta_\nu$ have a mild effect on the shape of the B clonal concentration (see 
Fig. \ref{fig:kappa}), while Fig. \ref{fig:r_pim} (top panel) shows that 
$\pi_\nu^-$ affects the decay of B cells after the infection peak, 
potentially 
sustaining long-term memory. Fig. \ref{fig:r_pim} (bottom panel) 
shows that although faster replicating antigens attain 
much higher concentrations, they are still removed, 
for $\epsilon\!>\!2r_\nu/(\kappa_\nu\pi_\nu^+(\infty))$,
on similar timescales, 
by triggering a stronger immune response.
Finally, Fig. \ref{fig:e} shows that at the critical value 
$\epsilon\!=\!2r_\nu/(\kappa_\nu \pi_\nu^+(\infty))$
the model predicts a very slight increase in the B cell population and successful clearance of the 
antigen (top panels), however stochastic fluctuations 
are anticipated to become important 
at criticality, as Gillespie simulations in Sec. \ref{sec:Gillespie} will 
confirm.   
Conversely, as soon as $\epsilon$ is lowered below zero, 
the system becomes unresponsive and is not able to fight a single antigen, 
even if replicating slowly (Fig. \ref{fig:e}, bottom panels). 
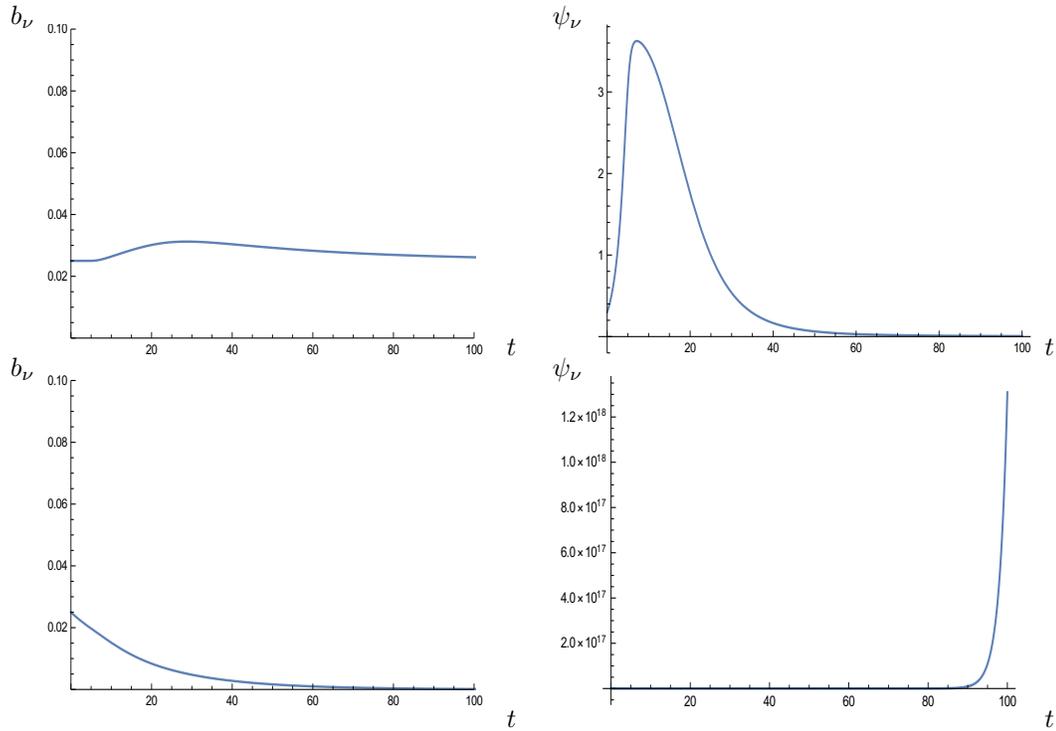
\begin{figure}[t!]
\hspace*{2cm}
\setlength{\unitlength}{0.55mm}
\begin{picture}(200,170)
\put(0,165){$b_\nu$}
\put(131,165){$\psi_\nu$}
\put(120,85){$t$}
\put(250,85){$t$}
\put(0,80){$b_\nu$}
\put(131,80){$\psi_\nu$}
\put(120,-5){$t$}
\put(250,-5){$t$}
\put(9,85){\includegraphics[width=105\unitlength,height=80\unitlength]{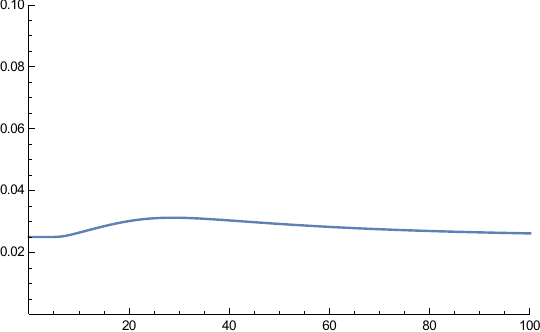}}
\put(142,85){\includegraphics[width=105\unitlength,height=80\unitlength]{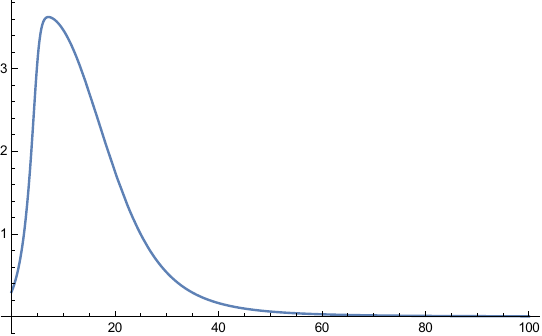}}
\put(9,0){\includegraphics[width=105\unitlength,height=80\unitlength]{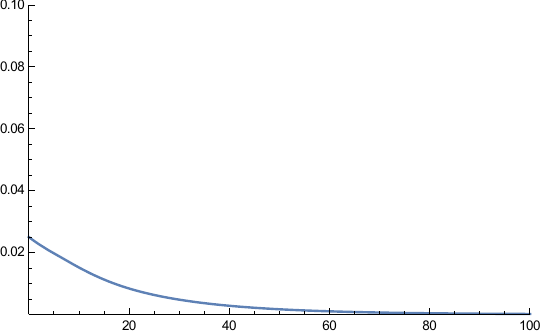}}
\put(133,0){\includegraphics[width=110\unitlength,height=80\unitlength]{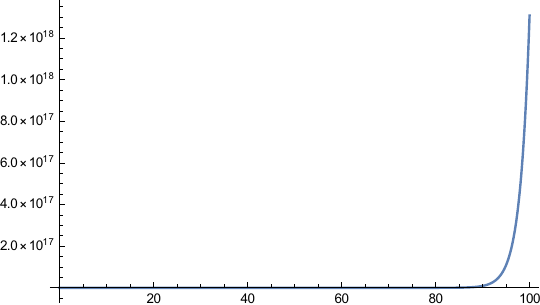}}
\end{picture}
\caption{Time evolution of B clonal density $b_\nu$ (left) and antigen concentration $\psi_\nu$
for $\beta=1$, $\pi_\nu^+=10\!\times\![\tanh(10)+\tanh(t-5)]$, $\pi_\nu^-\!=\!1$,
$\lambda_\nu\!=\!5$, $\delta_\nu\!=\!1$, $r_\nu\!=\!0.5$.
In the top panels $\epsilon\!=\!0.01$ (coincding with the critical value $2r_\nu/(\kappa_\nu \pi_\nu^+(\infty))$), while in bottom panels $\epsilon\!=\!-0.01$.
Initial conditions were chosen as $\psi(0)\!=\!0.3$ and $b_\nu(0)\!=\!\kappa_\nu |\epsilon|/2$. 
Plots show a very slight increase in the B cell density and successful clearance of the 
antigen at criticality, conversely, as soon as $\epsilon$ is lowered below zero, 
the system is unable to remove the antigen.
} 
\label{fig:e}
\end{figure}

These results suggest that for T-helper/T-suppressor ratios
$R>1$ (i.e. $\epsilon>0$) the host manages to remove 
completely the antigen, provided the affinity grows larger than an $\epsilon$-dependent value $\pi_\nu^+>2r_\nu/(\epsilon \kappa_\nu)$,   
while for $R<1$ 
the immune system is impaired 
and does not respond to the antigen, which replicates 
indefinitely in the host, no matter how large and fast the affinity $\pi_\nu^+$ between BCR and antigen grows.
From a more general point of view, the model captures the 
effectiveness and robustness of the immune system dynamics after exposure to an antigen and
predicts that the most important single 
parameter in determining whether the system is in a healthy or in an 
immuno-suppressed phase is $\epsilon$, directly related to the T-helper/T-suppressor ratio 
$R$, and in line with recent experiments 
\cite{Israel, ratio_cancer, ratio_HIV_review, 
ratio_inflamatory}. The model predicts the ratio to affect directly 
the location of the peak of the time-dependent antigen 
concentration, occurring 
when the affinity $\pi_\nu^+$ 
becomes larger than the $\epsilon$-dependent value 
$2r_\nu/(\epsilon \kappa_\nu)$, as well as 
the resting sizes of B clones, related to 
$\epsilon$ via $b_\nu=\epsilon \kappa_\nu/2~\forall~\nu$. 
This result is consistent with the equilibrium statistical mechanical analysis 
carried out in \cite{Alex_fast}, 
where the B cells density was shown to be a decreasing function of the 
T-suppressor cells density, at equilibrium. 
Note that $B_\nu=c_\nu N^{1-\gamma} b_\nu$ so the equilibrium size of B clone $\nu$ will depend on the 
number $c_\nu N^{1-\gamma}$ of T cells signaling B clone $\nu$. 
B clones which can be signaled by different T clones, will have 
higher values of $c_\nu$ and will therefore reach larger sizes at stationarity.

Finally we note that, although unstable, $b_\nu=0$ is always a fixed point of 
(\ref{eq:dyn_sys3}), 
even for $\epsilon>0$ and may be selected by means of stochastic fluctuations, 
leading to clonotypes extinction. 
Stochastic fluctuations are predicted to become relevant in the regime 
$\gamma=1$, where the number of B clones 
is extensive and the number of cells per clonotype is 
$\order{(N^0)}$, a scenario which 
has been observed 
in mouse \cite{TCR}.

\section{Simulations}
\label{sec:Gillespie}
In this section we employ the Gillespie algorithm \cite{Gillespie1977}, 
to simulate the biochemical reactions defined in Sec. \ref{sec:bio} 
at the single cell level, in the presence of a single 
antigen population. The underlying 
assumption of the algorithm is that cell populations
are well-mixed and 
interact in a finite volume. Since we do not consider antigen 
mutations and cross-reactivity effects between B clones and antigen, 
we will only consider one B clone and the ensemble of its conjugate T cells, so that we drop clonal indeces from now on.
\begin{figure}
\begin{picture}(200,330)
\put(0,-50){\includegraphics[clip, trim=0.5cm 4.8cm 0.5cm 4.8cm, width=\textwidth]{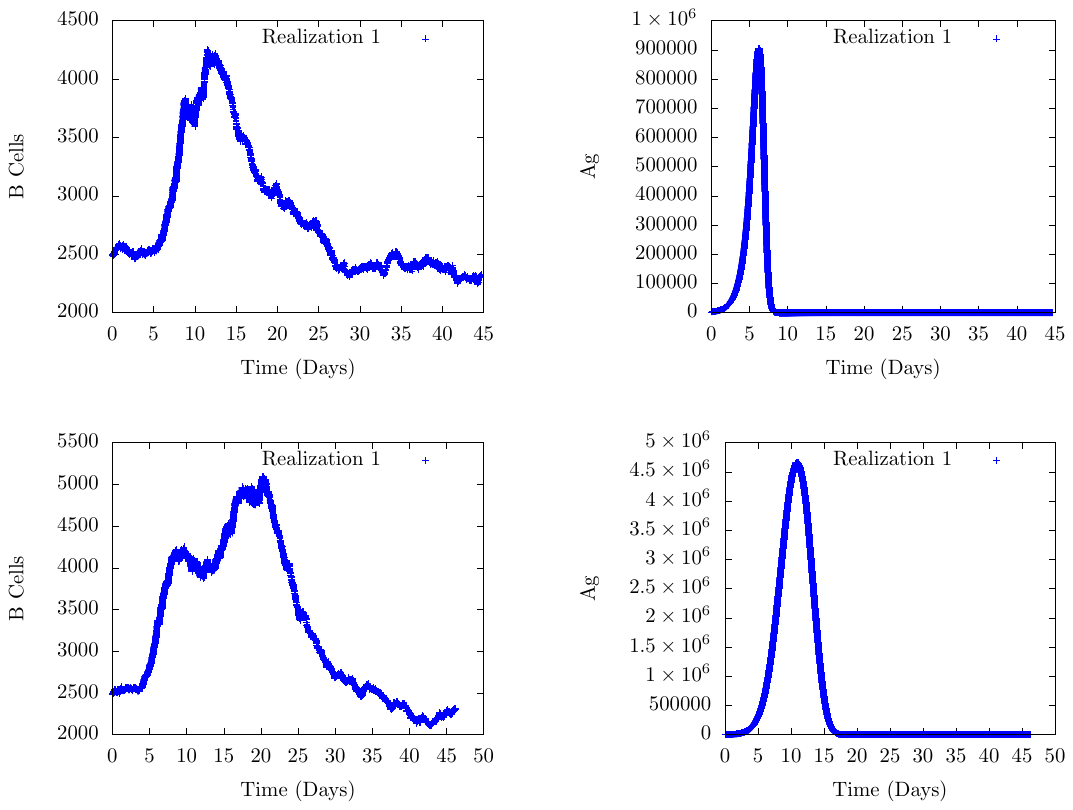}}
\end{picture}
\caption{Time evolution of B cell (left) and antigen population (right) for $r\!=\!1$, $\epsilon\!=\!0.5$, $\lambda\!=\!\delta\!=\!\pi^-\!=\!\beta\!=\!1$ and initial number of antigen cells 
$0.3\times 10^4$. Top panels: $\pi^+=10\!\times\![\tanh(10)+\tanh(t-7)]$. Bottom panels: $\pi^+=10\!\times\![\tanh(2.01)+\tanh[(t-20)/10]]$. Note that plots have different scales.}
\label{fig:Figure3}
\end{figure}
We have seven different species:
\begin{itemize}
\item B cells (B)
\item  Antigen presenting B cells (APB) 
\item Active and inactive suppressor T cells (S/S*)
\item Active and inactive helper T cells (H/H*)
\item Antigen (Ag)
\end{itemize}
\vspace*{2mm}
and ten different reactions:
\begin{itemize}
\item Antigen replication at rate $r$ (\ref{eq:Ag_r})
\item Binding between antigen and B cell, at rate $\pi^+$ (\ref{eq:APB})
\item Activation and deactivation of T cells, at rate $W=W_t(0)$ and $W'=W_t(1)$, respectively (see (\ref{eq:Wt}))
\bea
H &\rightarrow^{\!\!\!\!\!\! W}& H^\star
\\
S &\rightarrow^{\!\!\!\!\!\! W}& S^\star
\\
H^\star &\rightarrow^{\!\!\!\!\!\!\! W'}& H
\\
S^\star &\rightarrow^{\!\!\!\!\!\!\! W'}& S
\eea
We note that the activation of T cells due to binding with APB cells, is modelled here as a single-specie reaction 
with APB-dependent rate. Alternatively, one could model it as a two-specie reaction (\ref{eq:T_active}) with 
APB-independent rate, and account for spontaneous activation of T cells, due to noise, separately, as a single-specie reaction with constant rates. We have 
checked that the two implementations are equivalent, however the above implementation  
has the advantage of reducing the number of reactions and connects more explicitely with the equations of Sec. \ref{sec:model}, meaning that the same reaction rates can be used.
\item Unbinding of B cells and antigen, at rate $\pi^-$ (\ref{eq:APB_remove})
\item Expansion and contraction of B cells
\bea
H^\star+B &\rightarrow^{\!\!\!\!\!\! \lambda}& H^\star+2B
\\
S^\star +B &\rightarrow^{\!\!\!\!\!\! \lambda}& S^\star
\eea
\item B cells competition, at rate $\delta$ (\ref{eq:B_compete})
\end{itemize}
\vspace*{2mm}
\begin{figure}
\begin{picture}(200,330)
\put(0,-50){\includegraphics[clip, trim=0.5cm 5cm 0.5cm 5cm, width=\textwidth]{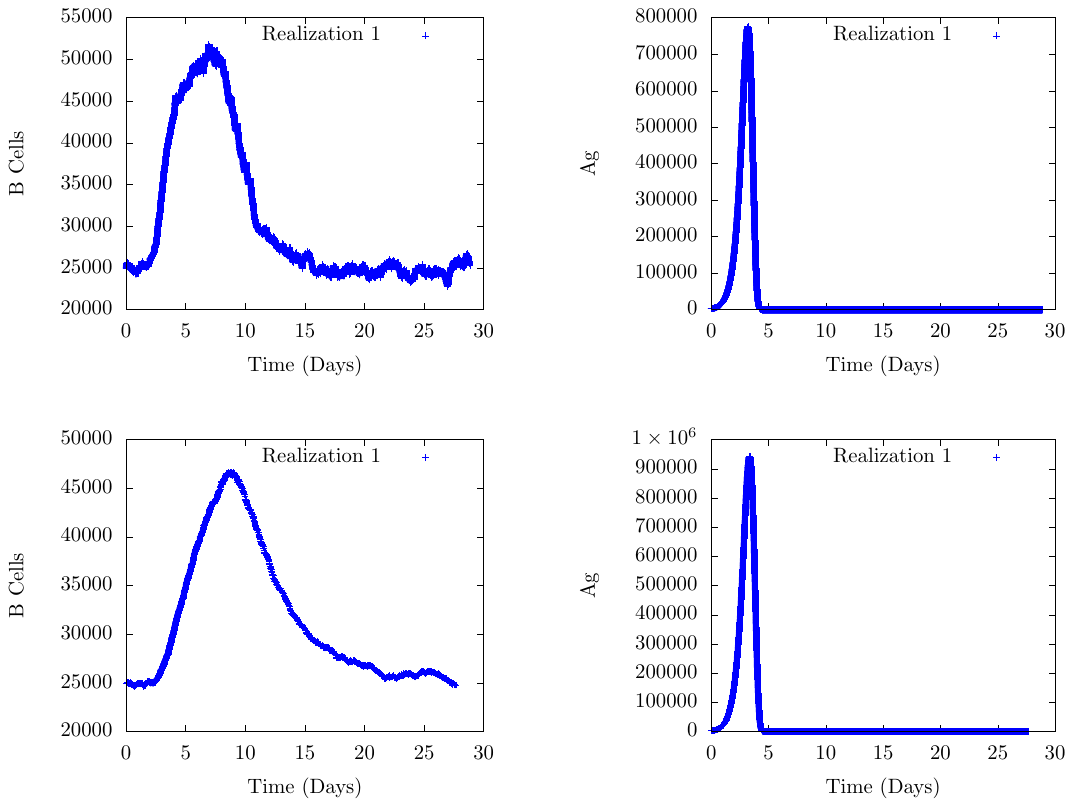}}
\end{picture}
\caption{Time evolution of B cell and antigen population respectively for $r=2$, $\epsilon =0.5$, $\pi^-=\beta=1$ and $\pi^+=10\times[\tanh(10)+\tanh(t-5)]$. Initial conditions as described in Figure \ref{fig:Figure3}. Top panels: $\lambda=10, \delta=1$. Bottom panels: $\lambda=1, \delta=0.1$. Note that the plots have different scales.}
\label{fig:Figure4}
\end{figure}
The initial conditions of the model are chosen from a well-mixed system in equilibrium at inverse noise level $\beta=1$, 
in the absence of antigen. 
The number of cells in the T clone
$cN^{1-\gamma}$ is set to $10^4$, of which
there are active and inactive T-helper and T-suppressor cells. 
Taken together, the initial total T-helper and total T-suppressor
populations, $N_H = H +{H^\star}$ and $N_S = S + {S^\star}$, are given by
$
N_H=cN^{1-\gamma}(1+\epsilon)/2$ and 
$N_S=cN^{1-\gamma}(1-\epsilon)/2$.
To find the ratio of active to inactive T cells, we refer 
to the update rule (\ref{eq:sigma}).
Given that $\beta=1$ and that the APB concentration is initially zero, 
we have that half of T cells are
initially active and half inactive. This approach of deriving 
initial T cell populations is equivalent
to finding the steady state solution to the ODE (\ref{eq:m_mu}) for $m_\mu$ 
and equating it to equation (\ref{eq:mmu}).
Finally, the steady state value of B cells is obtained from equation (\ref{eq:b_dyn}), by setting $m_\mu=\epsilon/2$, 
which gives
$
B= cN^{1-\gamma}\lambda \epsilon/(2 \delta)$.
We denote with ${\bf n}=(B,APB,S,S^\star,H,H^\star,{\rm Ag})$ 
the population vector associated to the seven different species
and we let ${\bf k}=(r,\pi^+, W, W, W', W', \pi^-, 
\lambda,\lambda,\delta)$ be the vector of the deterministic
reaction rates.  
Deterministic reaction rates ${\bf k}$ from the ODE approach 
can be converted to stochastic reaction rates ${\bf c}$, following 
a simple rule \cite{Gillespie1977}: $c_s = k_s$ if reaction $s$ 
involves a single reactant, 
$c_s = k_s/V$ for two distinct reactants, and $c_s = k_s/V^{n-1}$
for $n$ distinct reactants, where $V$ is the volume used to convert 
concentrations in the ODE approach to cell numbers in the stochastic 
simulation. 
This covers most possibilities except for when 
there are two reactants of the same species, in which case $c_s = 2k_s/V$, 
to account for the combinatorics of
reactions involving the same species.
%
%
%
\begin{figure}
\begin{picture}(200,330)
\put(0,-50){\includegraphics[clip, trim=0.5cm 5cm 0.5cm 5cm, width=\textwidth]{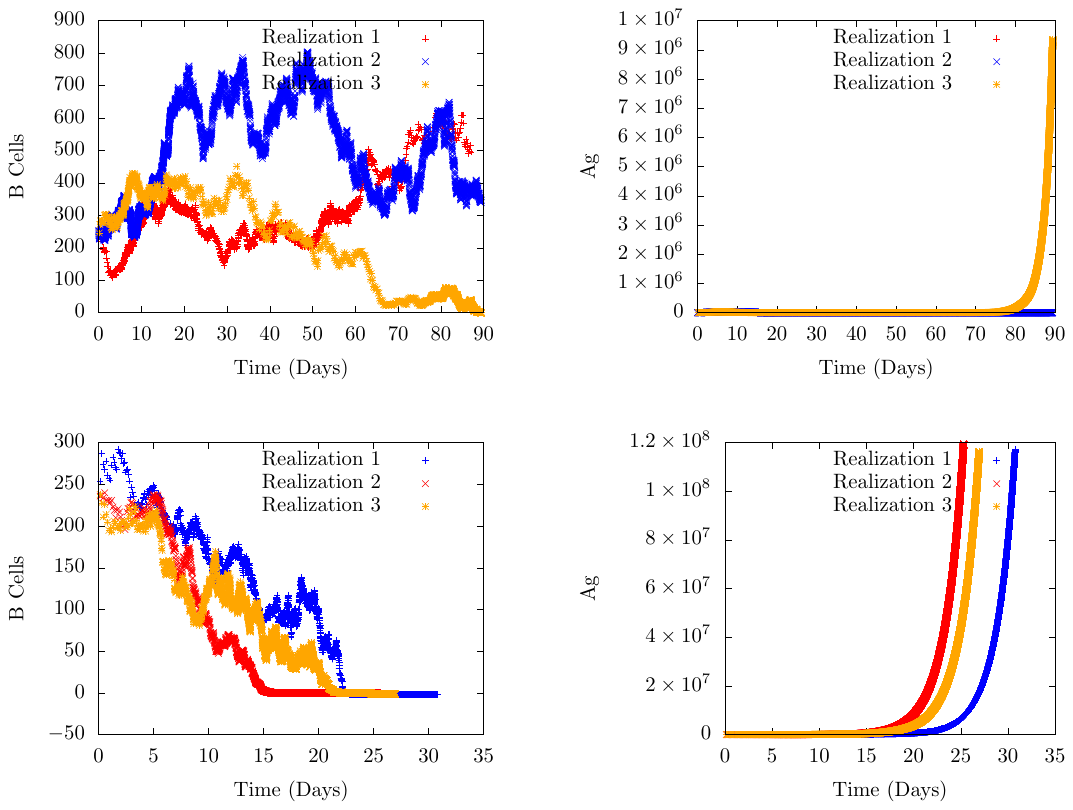}}
\end{picture}
\caption{Time evolution of B cell (left) and antigen population (right) for $r\!=\!0.5$, $\pi^+\!=\!10\!\times\![\tanh(10)+\tanh(t-7)]$, $\lambda\!=\!5$, 
$\delta\!=\!\pi^-\!=\!\beta\!=\!1$. 
Initial conditions are chosen as in Figure \ref{fig:Figure3}. Top Panels: 
$\epsilon\!=\!0.01$. Bottom panels: 
$\epsilon\!=\!-0.01$.}
\label{fig:Figure6}
\end{figure}
Since concentrations in the ODE approach were defined in terms of clonal sizes 
over the average number of conjugate T cells $cN^{1-\gamma}$ 
(as opposed to volume $V$)
we divide the deterministic reaction rates for reactions involving 
two species by $cN^{1-\gamma}$ and for reactions with the same single 
reactant appearing twice 
by $cN^{1-\gamma}/2$.
This leads to ${\bf c} =(r,\pi^+/(cN^{1-\gamma}), W, W, W', W', \pi^-, 
\lambda,\lambda,2\delta/(cN^{1-\gamma}))$.
%
At each iteration, the algorithm calculates $a_0 =\sum_{\ell=1}^{10} a_\ell$, where 
$a_{\ell}({\bf n}) = c_{\ell}({\bf n})h_{\ell}({\bf n})$ denotes the propensity and 
$h_\ell({\bf n})$ the distinct combinations of reactants in reaction $\ell$, 
and it determines the next reaction $s$ to execute as the one 
satisfying 
$\sum_{\ell=1}^{s-1}a_\ell<r_1 a_0\leq \sum_{\ell=1}^s a_\ell$
for a random number $r_1$ generated uniformly in $(0,1)$, and the time until it occurs via
$\tau=a_0^{-1}\ln r_2^{-1}$
for a random number $r_2$ generated uniformly in $(0, 1)$, hence it updates the number of molecules \cite{Gillespie1977}.

The results of Gillespie simulations are shown 
for B cells and antigen populations, 
in Figures \ref{fig:Figure3}, \ref{fig:Figure4} 
and \ref{fig:Figure6}. 
They are in agreement with those presented in Sec. 
\ref{sec:model}, confirming the 
validity of the reduced (mean-field) description of the system in terms of 
four ODEs, involving the macroscopic variable $m_\nu$, 
presented in Sec. \ref{sec:model}. 
In particular, all the simulations are seen to converge to the predicted steady states, within 
finite size fluctuations $\order{(N^{(1-\gamma)/2})}$.  
Modifying the time-dependence of $\pi^+$ shifts the peak and affects 
the maximum value of the antigen population 
(Fig. \ref{fig:Figure3}), increasing $\lambda$ or decreasing $\delta$ 
increases the population of B cells but it leads  
to the same qualitative dynamics of antigen removal following B clonal 
expansion (Fig. \ref{fig:Figure4}), 
and lowering $\epsilon$ below zero, the B cell population becomes extinct and 
the antigen population diverges (Fig. \ref{fig:Figure6}, bottom panels), 
consistently with the ODE approach. 
Conversely, the top panels of Fig. 
\ref{fig:Figure6} show that at criticality, 
different realizations show different behaviours, consistently with 
the expectation that fluctuations about 
the mean-field solution become important.
This suggests that stochasticity in the biological processes and 
discreteness of cells may drive immune systems operating near criticality 
in the suppressed phase.

\section{Model extensions}
\label{sec:extensions}
In this section we look at two extensions of the model, one 
allowing different rates for clonal expansion and suppression,
and the other inluding cross-reactivity effects, which occur when a BCR $\mu$
directed against antigen $\mu$ is also successful in binding with another antigen $\nu \neq \mu$.

\subsection{The role of clonal expansion and suppression rates}
In the model defined in previous sections, 
we made the assumption $\lambda_\nu^+= \lambda_\nu^-$. 
Allowing different kinetic coefficients for clonal expansion and suppression
$\lambda_\nu^+\neq \lambda_\nu^-$, 
modifies (\ref{eq:b_dyn}) to 
\be
\frac{d}{dt} b_\nu\!=\! 
b_\nu\left(\frac{\lambda_\nu^+ +\lambda_\nu^-}{2} m_\nu +\frac{\lambda_\nu^+-\lambda_\nu^-}{2} t_\nu
-\delta_\nu  b_\nu
\right)
\label{eq:b_2}
\ee
where $t_\nu=\bra t_\nu(\bsigma)\ket$ and 
\be
t_\nu(\bsigma)=\frac{1}{cN^{1-\gamma}}\sum_{i=1}^N \sigma_i \xi_i^\nu
\label{eq:tmu}
\ee
represents the density of activated T cells, regardless of their being helper 
or suppressors. Under the assumption of independence between 
$\eta$ and $\bxi$, that was used to derive (\ref{eq:m_many_ag}),
one has from (\ref{eq:mmu}) and (\ref{eq:tmu})
$m_\nu=\epsilon t_\nu$. Substituting in (\ref{eq:b_2}), we get
\bea
\frac{d}{dt} b_\nu&=& 
b_\nu\left(\tilde{\lambda}_\nu t_\nu -\delta_\nu  b_\nu
\right)
\label{eq:b_dyn2}
\eea
with
\be
\tilde{\lambda}_\nu=\frac{\lambda_\nu^+ +\lambda_\nu^-}{2} \epsilon +\frac{\lambda_\nu^+-\lambda_\nu^-}{2}
\label{eq:new_lambda}
\ee
Since $t_\nu \geq 0$ at all times, from (\ref{eq:b_dyn2}) we have that 
in the long-time limit $b_\nu=t_\nu \tilde{\lambda}_\nu/\delta_\nu $
for $\tilde{\lambda}_\nu>0$ and $b_\nu=0$ for $\tilde{\lambda}_\nu\leq 0$, where 
$t_\nu$ is the stationary solution of 
$$
\frac{d}{dt}t_\nu=\frac{1}{2}\left[1+\tanh\frac{\beta}{2}p_\nu\right]-t_\nu
$$ 
Then, from (\ref{eq:dyn_sys2}), it follows that 
the antigen is cleared 
for $r_\nu<\pi_\nu^+(\infty) \tilde{\lambda}_\nu/(2\delta_\nu)$, i.e. for
$$
\epsilon> \left(\frac{2r_\nu \delta_\nu}{\pi_\nu^+(\infty)}-\frac{\lambda_\nu^+-\lambda_\nu^-}{2}\right)\frac{2}{\lambda_\nu^+ +\lambda_\nu^-}
$$ 
Under the assumption that $\pi_\nu^+$ increases to values $\order{(10^2)}/{\rm 
day}$, 
while the other kinetic parameters are $\order{(1)}$,  
the critical value of $\epsilon$ above which the antigen is removed 
from the system is approximately given by 
$$
\epsilon\simeq -\frac{\lambda^+_\nu-\lambda_\nu^-}{\lambda_\nu^+ +\lambda_\nu^-}
$$
Recent experimental results suggesting that the 
onset of immunosuppression is associated with lymphocyte ratios 
close to one, i.e. 
$\epsilon\simeq 0$, justify a posteriori the assumption 
$\lambda_\nu^+\simeq \lambda_\nu^-$. 

\subsection{The role of cross-reactivity between B cells and antigens}
\label{sec:cross-r}
Our analysis has so far restricted to single epitope antigens, however, antigens have normally several epitopes which may be recognized by different B clones, leading to cross-reactivity effects. Here, we discuss briefly how these can be incorporated in the model. 
We introduce a variable $A_{\mu\nu}$ which takes value $1$ if BCR $\mu$ can bind antigen $\nu$ and $0$ otherwise. 
The model studied ealier, with no B-Ag cross-reactivity, 
corresponds to $A_{\mu\nu}=\delta_{\mu\nu}$, where BCR $\mu$ can only recognize 
antigen $\mu$.
In the presence of cross-reactivity, when an antigen $\nu$ enters the system, 
all the clones $\mu$ such that $A_{\mu\nu}=1$, will respond, so one has
\bea
\frac{d \psi_\nu}{dt}&=&\psi_\nu(r_\nu-\sum_\mu A_{\mu\nu}\pi_\mu^+  b_\mu)
\\
\frac{d p_\mu}{dt}&=&A_{\mu\nu}\pi_\mu^+ \psi_\nu b_\mu-\pi_\mu^- p_\mu
\\
\frac{dm_\mu}{dt}&=&\frac{\epsilon}{2}\!\left[1\!+\!\tanh 
\frac{\beta}{2}\left(p_\mu +\sum_{\rho\neq \mu} \xi^\rho p_\rho\right)  \!\right]
-m_\mu
\eea
with each B clone $\mu$ evolving according to equation (\ref{eq:b_dyn}).
We model the interactions $\{A_{\mu\nu}\}$ between B clones and antigen $\nu$, 
as random variables with distribution
$$
p(A_{1\nu},\ldots,A_{P\nu})=\delta_{A_{\nu\nu},1}\prod_{\mu\neq \nu} \left[\frac{d_\nu-1}{N^\gamma}\delta_{A_{\mu\nu},1}+\left(1-\frac{d_\nu-1}{N^\gamma}\right)\delta_{A_{\mu\nu},0}\right]
$$
where $d_\nu=\bra \sum_\mu A_{\mu\nu}\ket$ is the average number of B clones reacting with antigen
$\nu$, which we assume 
$\order{(1)}$.
In the long-time limit, APB densities will approach the value $p_\mu=A_{\mu\nu}b_\mu \psi_\nu a_\mu$ and 
$m_\mu$ will evolve according to
\bea
\frac{dm_\mu}{dt}=\frac{\epsilon}{2}\!\left[1\!+\!\tanh 
\frac{\beta}{2}\left(A_{\mu\nu}b_\mu \psi_\nu a_\mu +\sum_{\rho\neq \mu} \xi^\rho A_{\rho\nu}b_\rho \psi_\nu a_\rho \right)  \!\right]-m_\mu
\label{eq:m_cross}
\eea
The sum on the right end side is due to clonal interference, 
now comprising two effects: cross-reactivity between 
B and T cells and cross-reactivity between B cells and antigens.
If both types of cross-reactive interactions ($\xi$'s and $A$'s) are diluted, 
as postulated here, the sum is
$\order{(N^{-\gamma})}$, as it consists of $\order{(N^\gamma)}$ terms, 
each of order $\order{(N^{-2\gamma})}$. 
For $N\to \infty$, this vanishes and 
all the clones which are able to bind antigen $\nu$, 
of which there are $d_\nu$,
will expand via (\ref{eq:b_dyn}) and 
\bea
\frac{dm_\mu}{dt}=\frac{\epsilon}{2}\!\left[1\!+\!\tanh 
\frac{\beta}{2}\left(b_\mu \psi_\nu a_\mu \right)  \!\right]-m_\mu \quad \quad \forall~\mu: A_{\mu\nu}=1
\eea
Hence, cross-reactivity between B cells and antigens leads to the signaling 
of multiple B clones in parallel, but as long as cross-reactive 
interctions are diluted, they do not lead to signal interference.
On the other hand, since multiple B clones now
jointly contribute to the clearance of the antigen, 
the latter can be removed at smaller values of the affinities $\{\pi_\mu^+\}$, 
than those required in the absence of cross-reactivity.

Finally, we note that in the presence of B-Ag cross-reactivity,
interclonal competition may arise, as cross-reactive 
B clones may compete for the same resources
during clonal expansion, 
leading to coupled B clones dynamics 
$\dot b_\mu=b_\mu(\lambda_\mu m_\mu-\delta_\mu \sum_\rho A_{\mu \rho}b_\rho)$.
Although a full analysis of cross-reactivity effects goes beyond the scope 
of this work, 
the above discussion suggests that cross-reactivity may allow expansion of multiple clones in parallel and 
antigen clearence at smaller values of the affinity, 
on the other hand interclonal competition may arise and couple the dynamics of different clones, potentially
resulting in time-variation
of clones which cannot bind to the invading antigen directly.

\section{Conclusions}
\label{sec:conclusions}
In this work, we introduced a statistical mechanical model for the adpative immune system, 
which comprises B cells, T-helper cells, T-suppressor cells and antigens. 
When the ratio between T-helper and T-suppressor cells is above one,  
the model produces an immune response which is a unimodal function of time, 
in qualitative agreement with experimental 
observations, accomplishing, in particular, 
a complete removal of the antigen as well as lymphocyte homeostasis, 
for any chosen value of the control parameters 
(replication rates, cell death rates and noise level) 
and any chosen 
increasing function of time for the affinity between B cells and antigens, 
provided it increases beyond a critical value.
The model correctly predicts the existence of a lag time between 
infection and immune response detection, 
and informs on the role played by the T-helper/T-suppressor ratio
and by different kinetics parameters,  
on the steady state and relevant timescales (e.g. antigen density peak and B clonal contraction),
which could be potentially useful for parameters inference. 

As the ratio between T-helper and T-suppressor cells is tuned over a narrow region above one, 
the model exhibit a transition from a functional to an 
impaired phase, thus supporting the validity of the T-helper/T-suppressor ratio as an index of 
immuno-suppression. In the transitional state between the functional and the impaired state, noise plays an important 
role and B clones fail to contract after the antigen concentration reaches its peak, a pattern which is observed 
in ageing \cite{deborah}.

The identification of reliable markers of immuno-suppression is currently an active research field and several
indices have been correlated to disease prognosis in recent years  
(e.g. CD4+/Treg, Treg/CD8+, Th17/Treg, CD4+/CD4+CD25+, CD4+/CD8+ 
ratios, absolute 
numbers, and differences in cell numbers, of different sub-populations). With more lymphocyte sub-populations being uncovered, it becomes increasingly important to have 
models which are able to discern relevant parameters from accidental correlations.
The T-helper/T-suppressor ratio $R$ considered in this work, is given by the number of T-helper cells, 
i.e. CD4+ cells which are not regulatory, 
divided by the total number of CD8+ and T-regulatory cells, so 
$R=({\rm CD4}-{\rm Treg})/({\rm CD8}+{\rm Treg})$. This can be expressed in terms of the ratios 
$R_1$=CD4+/CD8+ and $R_2$=Treg/CD4+ commonly reported in the literature, as 
$R=R_1(1-R_2)/(1+R_1R_2)$. The latter is an increasing 
function of $R_1$ and a decreasing function of $R_2$, consistently with the experimental finding that
high values of the CD4+/CD8+ and the CD4+/Treg ratios both correlate with positive outcomes. This puts on 
a firmer ground the use of the CD4+/CD8+ and CD4+/Treg ratios for prognosis monitoring.
We argue, however, that given Treg cells are a small percentage of CD4+ cells, they may be subject 
to larger fluctuations, so the CD4+/CD8+ ratio
may be a more reliable index of immuno-suppression than CD4+/Treg. Also, we propose that the combination 
$R$ of the two may give further insights, especially when the ratio
Treg/CD4+ is abnormally large, as in cancer, autoimmunity and HIV diseases, so that $R$ may deviate 
significantly from $R_1$. 
We stress, however, that spatial heterogeneities have not been taken into account in the model and may play 
an important role. In particular, cell concentrations and kinetics may vary considerably across different tissues 
and organs, so the main merit of this approach lies  
in providing a mathematical tool to assess the relevance 
of the different cell ratios which have recently 
attracted experimental attention, 
rather than in determining accurately the values at which 
immune impairment occurs, which may be subject to spatial fluctuations.

Possible pathways for future research include the
effect of fast mutating antigens like cancerous cells and retrovirus, 
that manage to mutate before being removed from the system, 
and the mechanism by which HIV infection, cancer progression 
and immunosenescence alter the T-helper/T-suppressor ratio.
In addition, the biological realism of the model may be improved in several directions, 
by including interclonal competition and cross-reactivity effects, 
a more detailed modelling of the affinity maturation of B cells in the germinal centre, 
the role of self-antigens, T-T interactions and clonal expansion of active T cells. The latter process will lead to 
time variations of the number of T cells, which may be included 
in future extensions of the model by making the parameters 
$c_\mu$ dependent on time. The evolution of the size and diversity 
of T and B repertoires, encoded in the $c_\mu$ and $b_\mu$, 
in response to antigenic histories, may be an interesting pathway for 
future research.

We hope that even at this level of simplification, 
the model can offer a useful theoretical 
framework to understand homeostasis and impairment in the adaptive immunity, and 
complement recent experimental studies aimed at assessing the 
validity of the T-helper/T-suppressor ratio as a biological marker for 
immunosuppression.

\section*{Acknowledgements}
AA acknowledges the stimulating research environment provided by the 
EPSRC Centre for Doctoral Training in Cross-Disciplinary Approaches to 
Non-Equilibrium Systems (CANES) (EP/L015854/1) and Dr Alexander Mozeika for many interesting 
discussions.

\appendix

\section{Kramers-Moyal expansion of the Master equation}
\label{app:KM}
In this section, we use the master equation (\ref{eq:master}) 
for the evolution of the probability density $p_t(\bsigma)$, 
to derive dynamical equations
for the macroscopic parameters 
${\bf m}(\bsigma)=(m_1(\bsigma),\ldots,m_P(\bsigma))$. 
As a first step, we
derive an equation for the probability density 
${\mathcal P}(\bm)=\sum_{\bsigma}P(\bsigma)\delta(\bm -\bm(\bsigma))$
that the macroscopic parameters ${\bf m}(\bsigma)$ take values ${\bf m}$
\bea
\hspace*{-0.7cm}
\partial_t {\mathcal P}(\bm)
&=&\sum_\bsigma \delta(\bm-
\bm(\bsigma))
\sum_i [p_t(F_i \bsigma)W_t(1-\sigma_i)
-p_t(\bsigma)W_t(\sigma_i)]
\nonumber\\
\hspace*{-0.7cm}
&=&\sum_\bsigma 
\sum_i p_t(\bsigma)W_t(\sigma_i)[\delta(\bm-\bm(F_i\bsigma))
-
\delta(\bm-\bm(\bsigma))]
\label{eq:ME_m}
\eea
Next we work out the change $\Delta_{i\mu}(\bsigma)$
occurring in the parameter $m_\mu(\bsigma)$ when 
T cell $i$ is flipped
\bea
\Delta_{i\mu}(\bsigma)\!=\!m_\mu(F_i \bsigma)\!-\!m_\mu(\bsigma)
=\frac{1}{cN^{1-\gamma}}(1\!-\!2\sigma_i)\eta_i \xi_i^\mu
\eea
Carrying out a Kramers-Moyal expansion of 
(\ref{eq:ME_m}) in powers of $\Delta_{i\mu}(\bsigma)$ 
we obtain 
\bea
\hspace*{-0.7cm}
\partial_t {\mathcal P}(\bm)&=&\sum_\bsigma 
\sum_i p_t(\bsigma)W_t(\sigma_i)[-\sum_\mu \Delta_{i\mu}(\bsigma)\frac{\partial}{\partial m_\mu}\delta(\bm-\bm(\bsigma))
\nonumber\\
\hspace*{-0.7cm}
&&+\frac{1}{2}\sum_{\mu\nu}\Delta_{i\mu}(\bsigma)\Delta_{i\nu}(\bsigma)
\frac{\partial^2}{\partial m_\mu \partial m_\nu}
\delta(\bm-\bm(\bsigma))+\ldots]
\label{eq:ME_m2}
\eea
Next, we work out 
\bea
\hspace*{-0.7cm}
\sum_i W_t(\sigma_i)\Delta_{i\mu}(\bsigma)&=&
=\frac{1}{2cN^{1-\gamma}}\sum_i\left[1-2\sigma_i
+\tanh\frac{\beta}{2}\xi_i^\nu p_\nu\right]\eta_i \xi_i^\mu
\nonumber\\
\hspace*{-0.7cm}
&=&-m_\mu(\bsigma)+\frac{N^\gamma}{2c}\bra \eta \xi^\mu[1+\tanh\frac{\beta}{2}
\sum_\nu \xi^\nu p_\nu]\ket_{\eta,\bxi}
\label{eq:sumi}
\eea
where $\bra \cdot \ket_{\eta,\bxi}$ denotes the average over the distribution
$P(\eta,\bxi)$ defined in (\ref{eq:joint}).
Next we note that $p_\nu$ depends on $\bsigma$ only through $m_\nu(\bsigma)$, 
via (\ref{eq:p_dyn}) and (\ref{eq:b_dyn}).
Inserting (\ref{eq:sumi}) into (\ref{eq:ME_m2}), 
using the constraint $m_\mu(\bsigma)=m_\mu$ to remove the $\bsigma$-dependence 
of (\ref{eq:sumi}) and carrying out  
the summation over $\bsigma$, we obtain
\bea
&&\hspace*{-0.9cm}
\partial_t {\mathcal P}(\bm)\!=\!
-\sum_\mu\frac{\partial}{\partial m_\mu}[f_\mu(m_\mu,\bp){\mathcal P}(\bm)]
+\ldots
\label{eq:Liouville}
\eea
where 
\be
f_\mu(m_\mu,\bp)=-m_\mu+\frac{N^\gamma}{2c}\bra \eta \xi^\mu[1+\tanh\frac{\beta}{2}
\sum_\nu \xi^\nu p_\nu]\ket_{\eta,\bxi}
\ee
One can show that, away from criticality, higher order terms, arising from the second term in 
the square brackets of (\ref{eq:ME_m2}), are at most $\order{(N^{\gamma-1})}$,
hence for large $N$ and $\gamma<1$ they vanish. 
In this limit, (\ref{eq:Liouville}) becomes 
a Liouville equation, that corresponds to the
deterministic (or mean-field) equation (\ref{eq:av_m}) for the evolution of the variables $\bm(\bsigma)$.
For large but finite values of $N$, fluctuations of the stochastic parameters 
$m_\mu(\bsigma)$ about the deterministic values $m_\mu=\bra m_\mu(\bsigma)\ket$
will be $\order{(N^{(\gamma-1)/2})}$.

\end{document}